\documentclass[aps,pra,twocolumn,superscriptaddress,floatfix,footinbib,notitlepage,groupaddress,showpacs]{revtex4-1} 
\usepackage{times,color,amsfonts,amssymb,amsbsy,amsthm,amsmath,graphics,graphicx,bm,bbm,makecell,mathtools,dsfont}
\usepackage{upgreek}
\DeclareFontFamily{OT1}{pzc}{}
\DeclareFontShape{OT1}{pzc}{m}{it}%
              {<-> s * [1.25] pzcmi7t}{}
\DeclareMathAlphabet{\mathpzc}{OT1}{pzc}%
                                 {m}{it}
\let\oldsqrt\sqrt
\def\sqrt{\mathpalette\DHLhksqrt}
\def\DHLhksqrt#1#2{%
\setbox0=\hbox{$#1\oldsqrt{#2\,}$}\dimen0=\ht0
\advance\dimen0-0.2\ht0
\setbox2=\hbox{\vrule height\ht0 depth -\dimen0}%
{\box0\lower0.4pt\box2}}
\begin{document}

\title{Long-distance continuous-variable quantum key distribution with quantum scissors}

\author{Masoud Ghalaii}
\affiliation{Faculty of Engineering and Physical Sciences, University of Leeds, Leeds LS2 9JT, United Kingdom}
\author{Carlo Ottaviani}
\affiliation{Computer Science and York Centre for Quantum Technologies, University of York, York YO10 5GH, United Kingdom}
\author{Rupesh Kumar}
\affiliation{Department of Physics, University of York, York YO10 5DD, United Kingdom}
\author{Stefano Pirandola}
\affiliation{Computer Science and York Centre for Quantum Technologies, University of York, York YO10 5GH, United Kingdom}
\affiliation{Research Laboratory of Electronics, Massachusetts Institute of Technology (MIT), Cambridge, MA, USA}
\author{Mohsen Razavi}
\affiliation{Faculty of Engineering and Physical Sciences, University of Leeds, Leeds LS2 9JT, United Kingdom}

\begin{abstract}
We investigate the use of quantum scissors, as candidates for non-deterministic amplifiers, in continuous-variable quantum key distribution. Such devices rely on single-photon sources for their operation and as such, they do not necessarily preserve the Guassianity of the channel. Using exact analytical modeling for the system components, we bound the secret key generation rate for a protocol that uses quantum scissors. We find that, for certain non-zero values of excess noise, such a protocol can reach longer distances than the counterpart with no amplification. This sheds light into the prospect of using quantum scissors as an ingredient in continuous-variable quantum repeaters. 
\end{abstract}

\maketitle

%%%%%%%%%%%%%%%%%%%%%%%%%%%%%%%%%%%%%
\section{Introduction}
Quantum key distribution (QKD)
\cite{Pirandola:RevQKD2019,Bennett_BB84,Ekert_1991} addresses the problem of sharing secret keys between two users. 
Such keys can then be used for secure communications. While original QKD protocols \cite{Bennett_BB84,Ekert_1991,Gisin_QCrypRev_2002,Scarani_QKDRev_2009} rely on encoding classical bits of information in discrete quantum states, such as the polarization of single photons, one can also exploit continuous-variable QKD (CV\:QKD) protocols, where the bits are encoded on the quadratures of light \cite{Grosshans_GG02_PRL,Grosshans_GG02_Nature,Braunstein_QICVRev_2005,Weedbrook_GaussQIRev_2012}. In particular, the recent progress in CV\:QKD systems has placed them in a competitive position with their conventional discrete-variable counterparts \cite{Jouguet_LD_CVQKD_NatPhoton, Pirandola2015}. For instance, contrary to discrete-variable QKD protocols, which require single-photon detectors, CV\:QKD uses coherent measurement schemes, such as homodyne and/or heterodyne detection, to measure light quadratures, {compatible with high-rate coherent telecommunications systems}  \cite{Hirano_PulsedHOM,Yonezawa_QTele_Exp,Yokoyama_NatPhoton}. Moreover, CV\:QKD protocols can be the better choice over short distances than the metropolitan zones \cite{Pirandola2015}. Once it comes to long distances, however, CV\:QKD has its own challenges to compete with discrete-variable QKD \cite{Jouguet_LD_CVQKD_PRA}.
This paper examines how the security distance can be enhanced in CV\:QKD systems by using realistic non-deterministic amplification \cite{Blandino_CVQKD_idealNLA}. 

One of the proposed solutions to improve the rate-versus-distance performance of CV\:QKD protocols is to use noiseless linear amplifiers (NLAs) \cite{Blandino_CVQKD_idealNLA,Zhang_2Way_CVQKD_Amp}. It is known that deterministic amplification cannot be noise free \cite{Pandey_Qlimits_Amp}. An NLA can only then work {\em probabilistically}. This inevitably reduces the key rate by a factor corresponding to the success rate of the NLA, which implies that, at short distances, the use of NLAs may not be beneficial. The key rate may, however, increase at long distances because of the improvement in the signal to noise ratio. That is, while the number of data points we can use for key extraction is less, the quality of the remaining points could be also high, such that a larger number of secret key bits can be extracted.
This has been shown theoretically by treating the NLA as a probabilistic, but noiseless, black box, where an upper bound on success probability, $1/g^2$ with $g$ being the amplification gain, was used \cite{Blandino_CVQKD_idealNLA}.

The story can be quite different when we replace the above ideal NLA with realistic systems that offer NLA-like functionality. For instance, one of the most basic structures for an NLA is a quantum scissor (QS), which combines the incoming light with a single photon \cite{Pegg_QSs,Ralph_Lund_QSNLA}. While under weak signal assumptions, a QS can be approximated as an NLA, more precise analysis reveals that its operation is not necessarily noiseless. This is particularly important because in many CV\:QKD protocols the transmitted signal does not have a fixed intensity, and realistic NLAs often treat different input signals differently. This is more or less true for other proposals that implement the NLA operation \cite{Eleftheriadou_QAmp,Fiurasek,Xiang_NatPhys,Ferreyrol_ImpNLA_PRL,Donaldson_ImpNLA_PRL,Barbieri_NLA_Exp_Rev}.

In this paper, we provide a realistic account of what a QS can offer within a CV\:QKD setup. In particular, using an exact model for the QS setup, we analyze the secret key rate of a Gaussian modulated protocol, whose receiver unit is equipped with a QS. One of the implications of our exact modeling for the QS is that we cannot directly apply standard key rate calculation techniques that rely on the Gaussianity of the output states. This will make the exact calculation of the key rate cumbersome. We manage this problem by using relevant bounds for certain components of the key rate. We investigate the extent at which the use of quantum scissors can increase the security distance in CV QKD systems.

\begin{figure}[t]
	\centering
	\includegraphics[width=1\linewidth]{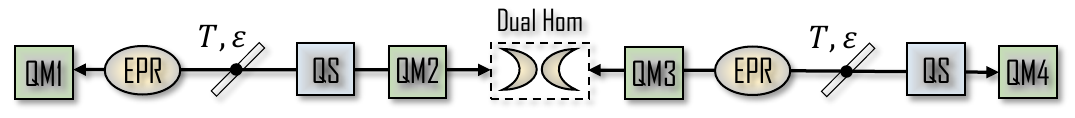}
	\caption{A two-leg quantum repeater module as proposed in   \cite{Dias_Ralph_CVQRs}. Each leg is composed of an EPR source generating two-mode squeezed vacuum states, a quantum scissor (QS), and two quantum memory (QM) units. Beam splitters with transmissivity $T$ characterize the loss in each leg, with excess noise represented by $\varepsilon$. Upon successful operation of the QS in each leg, the output of the QS and the EPR source are stored in respective quantum memories. When both legs are ready, a joint dual homodyne (Dual Hom) measurement is performed on the quantum states stored in QM2 and QM3, which swaps entanglement to QM1 and QM4. }
	\label{fig:QR}
\end{figure}

One of our key incentives for carrying out the above analysis is to provide insights into the applicability of other proposals for CV quantum repeaters \cite{Dias_Ralph_CVQRs,Furrer_Munro_CVQRs, Guha_CV_Repeater} for QKD operation. The QS-equipped CV QKD link that we consider here contains the elementary repeater (error correction) link used in the repeater setup of   \cite{Dias_Ralph_CVQRs}, and as such a poor performance for this basic building block could cast shadow on the usefulness of any larger quantum repeater setup that relies on such elementary links. In the repeater setup of \cite{Dias_Ralph_CVQRs}, CV teleportation is used to swap entanglement between already entangled links, represented by QM1-QM2 and QM3-QM4 in Fig.~\ref{fig:QR}. Each of such links have been entangled by sending one half of a two-mode squeezed vacuum state, represented by EPR boxes, through a lossy channel. The received signal will then be amplified, in a probabilistic way, by the QS module, and will be stored in the corresponding quantum memory (QM). Note that, considering the non-deterministic behavior of the QS, use of QM modules is necessary if we want to achieve any rate enhancement from our repeater setup. The dual homodyne module will then effectively perform entanglement swapping in the CV domain once both links have had successful QS operations.

Note that the above repeater setup must use a {\em physical} QS implementation, and not a virtual one, in order to offer any rate advantage. That is, the class of measurement-based NLA (MB-NLA) implementations \cite{Fiurasek_VirNLA,Walk_CVQKD_postsel, Chrzanowski_MBNLA}, which rely on data post-selection, would not be suitable for such CV repeaters. Due to reliance of MB-NLAs on classical post-selection, the state of QM2 and QM4 must effectively be measured before the entanglement swapping can be done. Even if we do not consider the applications of our considered setup in CV repeater settings, one must be cautious with typically poor success probability of MB-NLAs compared to that of physical NLAs \cite{Zhao_VirNLA}. This suggests that the use of physical NLAs in CV QKD systems is still of interest, and, in fact, one may favour a physical realization of an NLA over its virtual post-measurement implementation due to restrictions on the MB-NLA \cite{Bernu_NLA_EPR_dist}. Our work here would shed more light into the applicability of such physical realizations by offering an accurate analysis of the underlying system. 

The manuscript is structured as follows. 
In Sec.~\ref{sec:system_description}, we describe details of the proposed system.
In Sec.~\ref{sec:quantum_scissor}, by analyzing input-output characteristic functions of a single QS, we calculate the exact output state and success probability of the QS\:NLA in \cite{Ralph_Lund_QSNLA}. We also study the non-Gaussian behavior of this system. In Sec.~\ref{sec:secret_key_analysis}, we present the key rate analysis of the CV\:QKD link with a single QS as part of its receiver. In Sec.~\ref{sec:numerical_results}, we discuss the numerical results.
Finally, Sec.~\ref{sec:conclusion} concludes the paper. 

%----------------------------------------------------
%----------------------------------------------------

\section{System Description}
\label{sec:system_description}
\begin{figure}[t]
	\centering
	\includegraphics[width =0.8\columnwidth]{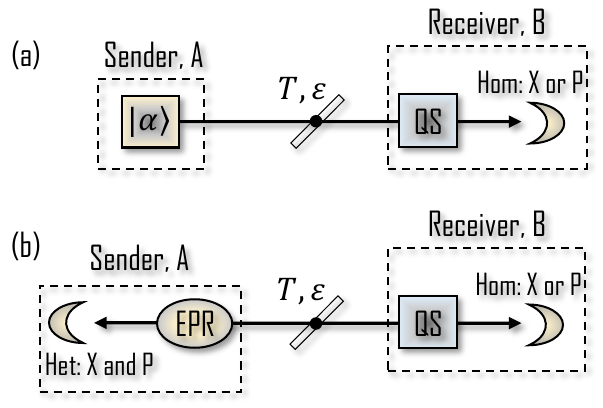} 
	\caption{(a) Schematic view of CV\:QKD link with an additional quantum scissor at the receiver. (b)  Entanglement-based CV\:QKD protocol equivalent to (a). Hom and Het represent, respectively, the homodyne detection and heterodyne detection modules.}
	\label{fig:cvqkd}
\end{figure}

In this section, we describe our proposed setup for the QS-amplified CV\:QKD protocol. We assume that the sender, Alice (A), is connected to the receiver, Bob (B), via a quantum channel; see Fig.~\ref{fig:cvqkd}(a). The protocol runs along the same lines as proposed by Grosshans and Grangier in 2002 (GG02) \cite{Grosshans_GG02_PRL,Grosshans_GG02_Nature,Lodewyck_25kmCVQKD,Kumar_CVQKDwithDWDM}. That is, in every round, Alice transmits a coherent state $|\alpha\rangle$ to Bob, where $\alpha = x_A +  i p_A$, with real parameters $x_A$ and $p_A$ being chosen randomly according to the following Gaussian probability density functions: 
\begin{align}
	\label{Eq:InputDist}
	f_{X_A}(x_A)=\frac{e^{-\frac{x_A^2}{V_A/2}}}{\sqrt{\pi V_A/2}}   ~~~ \text{and} ~~~  f_{P_A}(p_A) = \frac{e^{-\frac{p_A^2}{V_A/2}}}{\sqrt{\pi V_A/2}},
\end{align}
where $V_A$ is the modulation variance in the shot-noise units. At the receiver, however, we equip Bob with a single QS before the homodyne module used in GG02. Upon a successful QS operation, Bob randomly chooses to measure $\hat x_B = \hat a_B + \hat a_B^\dag$ or $\hat p_B = (\hat a_B - \hat a_B^\dag)/i$, where $\hat a_B$ represents the annihilation operator for the output mode of the QS. During the sifting stage, Bob would then publicly declare his measurement choices as well as the rounds in which the QS has been successful. Alternatively, one can use the equivalent entanglement-based (EB) scheme of Fig.~\ref{fig:cvqkd}(b), where Alice's source is replaced with an EPR source followed by heterodyne detection on one of the two modes of the state (by EPR source we mean a two-mode squeezed vacuum state \cite{Weedbrook_GaussQIRev_2012}). In either case, we assume that Bob can reconstruct, in an error-free way, the phase reference for the local oscillator used in his homodyne detection. By using post-processing techniques, Alice and Bob extract a key from the subset of data for which the QS has been successful.

%----------------------------------------------------
%----------------------------------------------------

\begin{figure}[t]
	\centering
	\includegraphics[width=0.6\columnwidth]{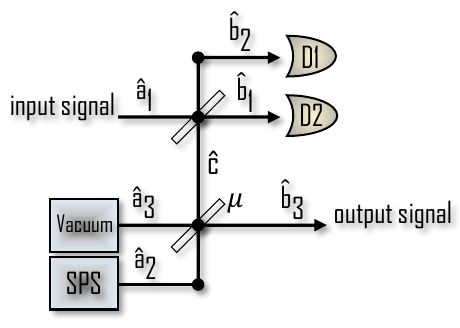}
	\caption{The schematic diagram of a quantum scissor. Here, we assume that an on-demand ideal single-photon source (SPS) is in use, and that the single-photon detectors have unity efficiencies.} 
	\label{fig:QS}
\end{figure}

Quantum scissors are the main building blocks in the NLA proposed by Ralph and Lund \cite{Ralph_Lund_QSNLA}. At the core of a QS, there is a partial Bell-state measurement (BSM) module, with a balanced beam splitter followed by two single-photon detectors, in the space spanned by number states $|0\rangle$ and $|1\rangle$. This BSM module is driven by an asymmetric entangled state $ |\psi\rangle = \sqrt{\mu}|1\rangle_{\hat{c}} |0\rangle_{\hat{b}_3} + \sqrt{1-\mu} |0\rangle_{\hat{c}} |1\rangle_{\hat{b}_3}$, generated by a single photon that goes through a beam splitter with transmittance $\mu$; see Fig.~\ref{fig:QS}. For an input state in the $|0\rangle$-$|1\rangle$ space, the QS could then offer an asymmetric teleportation functionality, whenever the BSM operation is successful, i.e., when only one of D1 or D2 detector in Fig.~\ref{fig:QS} clicks. For instance, in the particular case of a weak coherent state input $|\alpha\rangle_{\hat{a}_1} \approx |0\rangle_{\hat{a}_1} + \alpha |1\rangle_{\hat{a}_1} $, with $|\alpha| \ll 1$, a single click could come from the single-photon component in the entangled state $|\psi\rangle $ and/or the input state. In that case, the output state, after renormalization, can be approximated by $ |0\rangle_{\hat{b}_3} + \alpha g |1\rangle_{\hat{b}_3} \approx |\alpha g\rangle_{\hat{b}_3}$, for $|g\alpha| \ll 1$, where $g = \sqrt{(1-\mu)/\mu}$ represents the amplification gain of the QS. Under these assumptions, the success probability for the QS operation is given by $P_{\mathrm{succ}}^{\mathrm{RL}}(\alpha) \approx \mu +(1-\mu)|\alpha|^2$. 
Note that, in the above description, the essential assumption for a QS to possibly operate as an NLA is that $|\alpha|\ll 1$. 

There are two reservations in using the above asymptotic approach for analyzing a QS-based CV\:QKD system. First, note that the output state of a QS is always in the space spanned by single-photon and vacuum states. By approximating the output state as a coherent state, we are introducing some errors, which can affect the security of the system. More precisely, the transition from a coherent state to a single-photon state is a non-Gaussian one, whose effect must be carefully considered in the security analysis. Secondly, in the GG02 protocol, the coherent states are chosen randomly via Gaussian distributions; hence, the input states to the QS may not necessarily satisfy the assumption $|\alpha| \ll 1$. 

In order to resolve the above issues, in our work, we find the {\em exact} output state and probability of success for an arbitrary coherent state at the input of a QS. This will be detailed in Sec.~\ref{sec:quantum_scissor}.
We note that one can implement a QS/NLA which truncates input states to first $N$ Fock states \cite{McMahon_Optimal_QS,Jeffers_Two-photon_QS}. Here we limit ourselves to the single-photon truncation. 
We then apply our findings to the key rate analysis of a QS-equipped CV\:QKD system. For simplicity, we assume that the required single-photon source (SPS) in the QS is ideal and on-demand. Single-photon detector efficiencies are also assumed to be unity. Our analysis can, nevertheless, be extended to account for the imperfections in the source and detectors.

%------------------------------------------------------------------
%------------------------------------------------------------------

\section{Quantum scissors: input-output relationship}
\label{sec:quantum_scissor}
In this section, we first obtain an exact input-output relationship for a QS driven by a coherent state. We use characteristic functions to model the input and output states. For a joint, $M$-mode, state $\hat{\rho}$, where each mode $j$ is represented by an annihilation operator $\hat{a}_j$, the antinormally-ordered characteristic function is given by
\begin{align}
	\label{antinorm-func}
	\chi_{\mathrm{A}}^{\hat{\rho}} (\xi_1, \dots, \xi_M)= \Big\langle \bigotimes_{j=1}^M \hat{D}_{\mathrm{A}}(\hat{a}_j, \xi_j) \Big\rangle_{\hat{\rho}},
\end{align}
where $\langle \circ \rangle_{\hat{\rho}} \equiv \mathrm{tr}[\hat{\rho} \circ]$ and
$\hat{D}_{\mathrm{A}}(\hat{a}, \xi) = e^{-\xi^\ast\hat{a}} e^{\xi\hat{a}^{\dagger}}$ is the antinormally-ordered displacement operator with $\xi^\ast$ being the complex conjugate of the complex number $\xi=\xi_r + i \xi_i$, with $\xi_r$ and $\xi_i$ are real numbers.  
The density matrix $\hat{\rho}$ and its antinormally-ordered characteristic function are connected via a Fourier-transform as follows
\begin{align}
	\label{rec-state}
	\hat{\rho}= & \int \frac{d^2\xi_1}{\pi} \dots \int \frac{d^2\xi_M}{\pi} 
	\chi_{\mathrm{A}}^{\hat{\rho}} (\xi_1, \dots, \xi_M)
	\bigotimes_{j=1}^M \hat{D}_{\mathrm{N}}(\hat{b}_j, \xi_j) ,
\end{align}
where $\hat{D}_{\mathrm{N}}(\hat{a}, \xi) = e^{\xi\hat{a}^{\dagger}} e^{-\xi^\ast\hat{a}}$ is the normally-ordered displacement operator and $\int d^2\xi = \int_{-\infty}^{+\infty} d\xi_r \int_{-\infty}^{+\infty} d\xi_i$.

\begin{figure}[t]
	\centering
	\includegraphics[scale=0.75]{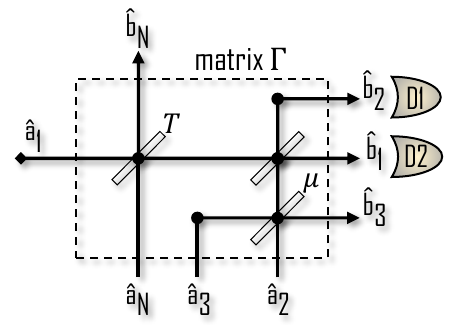}
	\caption{The quantum channel and the QS are considered as a combined system with input modes $\hat{a}_1 - \hat{a}_3$ and $\hat{a}_\mathrm{N}$ and output modes $\hat{b}_1 - \hat{b}_3$ and $\hat{b}_\mathrm{N}$. The transformation matrix of the system is given by \eqref{matrix-TQS}.}
	\label{fig:PandM}
\end{figure}

In the following, we use the above formulation to analyze the setup in Fig.~\ref{fig:PandM}, which includes a QS driven by an arbitrary coherent state through a lossy channel with transmissivity $T$ and excess noise $\varepsilon$. 

%------------------------------------------------------------
%------------------------------------------------------------

\subsection{Pre-measurement state}
For the setup in Fig.~\ref{fig:PandM}, we can use the well-known relationships for beam splitters to relate the four input modes to the four output modes. The dashed box $\Gamma$ is a linear optics circuit, for which such input-output relationships can be obtained. In particular, considering the input modes represented by $\mathcal{A}^T=[\hat{a}_1~ \hat{a}_2~ \hat{a}_3~ \hat{a}_\mathrm{N}]$ and output modes $\mathcal{B}^T=[\hat{b}_1 ~\hat{b}_2 ~\hat{b}_3 ~\hat{b}_\mathrm{N}]$, we find $\mathcal{B}=\Gamma \mathcal{A}$, where the transformation matrix
\begin{align}
	\label{matrix-TQS}
	\Gamma=  
	\left(\begin{array}{cccc}
		\sqrt{\frac{T}{2}} & \sqrt{\frac{\mu}{2}} & -\sqrt{\frac{1-\mu}{2}} & \sqrt{\frac{1-T}{2}}     \\
		-\sqrt{\frac{T}{2}}  & \sqrt{\frac{\mu}{2}}& -\sqrt{\frac{1-\mu}{2}} & -\sqrt{\frac{1-T}{2}}     \\
		0  & \sqrt{1-\mu}& \sqrt{\mu} & 0     \\
		-\sqrt{1-T} & 0 & 0 & \sqrt{T}    
	\end{array}\right) 
\end{align}
is a unitary orthogonal matrix, i.e., $\Gamma^\mathrm{T}= \Gamma^{-1}$.
The output antinormally-ordered characteristic function can then be expressed in terms of the input one by
\begin{align}
	\label{out-in-Anti}
	\chi_{\mathrm{A}}^{\mathrm{out}} (\xi_1,\xi_2,\xi_3,\xi_{\rm N}) = &  
	\big\langle \prod_{m=1}^3 \hat{D}_{\mathrm{A}}(\hat{b}_m, \xi_m) \hat{D}_{\mathrm{A}}(\hat{b}_{\rm N}, \xi_{\rm N})   \big\rangle   \nonumber \\
	= &  \big\langle  \prod_{m=1}^3  \hat{D}_{\mathrm{A}}(\hat{a}_m, \lambda_m) \hat{D}_{\mathrm{A}}(\hat{a}_{\rm N}, \lambda_{\rm N}) \big\rangle   \nonumber \\
	= & \chi_{\mathrm{A}}^{\mathrm{in}} (\lambda_1,\lambda_2,\lambda_3,\lambda_{\rm N}),
\end{align}
where $ [\lambda_1~ \lambda_2~ \lambda_3 ~\lambda_{\rm N}]^\mathrm{T} = \Gamma^\mathrm{T}  [\xi_1~ \xi_2~ \xi_3 ~\xi_{\rm N}]^\mathrm{T}$, with $\Gamma^\mathrm{T}$ being the transpose of $\Gamma$.
Here, we make use of the fact that  $\hat{D}_{\mathrm{A}}(s\hat{a}, \xi) = \hat{D}_{\mathrm{A}}(\hat{a}, s\xi), ~s\in \mathbbm{R}$, and $\langle \hat{D}_{\mathrm{A}}(\hat{a}, \xi_1)\hat{D}_{\mathrm{A}}(\hat{a}, \xi_2)\rangle = e^{\xi_1\xi^\ast_2} \langle \hat{D}_{\mathrm{A}}(\hat{a}, \xi_1 +\xi_2) \rangle$. 

Next, we consider the particular input state
\begin{align}
	\label{input_st_QS}
	\hat{\rho}_{\rm in} = &  |\alpha\rangle_{\hat{a}_1}\langle \alpha | \otimes |1\rangle_{\hat{a}_2} \langle 1 | \otimes |0\rangle_{\hat{a}_3}\langle 0 |  \otimes \int d^2\beta f_{\varepsilon}(\beta) |\beta \rangle_{\hat{a}_\mathrm{N}} \langle \beta |,
\end{align}
where $f_{\varepsilon}(\beta)=\frac{e^{-\frac{|\beta |^2}{\varepsilon/2}}}{\pi \varepsilon/2}$, with $\varepsilon$ being the channel excess noise. This corresponds to a Gaussian attack by Eve, enabled by an entangling cloner \cite{Navascues_EntanglingCloner}, which we later use in forthcoming sections.
For the above set of input states, the output characteristic function has the following expression 
\begin{align}
	\chi_{\mathrm{A}}^{\mathrm{out}} (\xi_1, \xi_2, \xi_3, \xi_{\rm N}) = &  \mathrm{tr} \big[ \hat{\rho}_{\rm in} \hat{D}_{\mathrm{A}}(\hat{a}_1, \lambda_1)  \hat{D}_{\mathrm{A}}(\hat{a}_2, \lambda_2) \nonumber \\
	& \hat{D}_{\mathrm{A}}(\hat{a}_3, \lambda_3) \hat{D}_{\mathrm{A}}(\hat{a}_{\rm N}, \lambda_{\rm N}) \big].
\end{align}
By using the transformation matrix $\Gamma$, this can be re-written as the following
\begin{align}
	\label{A-func-out}
	\chi_{\mathrm{A}}^{\mathrm{out}}(\xi_1,\xi_2,\xi_3 , \xi_\mathrm{N})= &  e^{-\frac{T}{2}|\xi_1- \xi_2 -\sqrt{2}\tau \xi_\mathrm{N}|^2 }  \nonumber \\
	& \times e^{\sqrt{2T} i  \text{Im}[\bar{\alpha} (\xi_1- \xi_2 -\sqrt{2}\tau \xi_\mathrm{N}) ] }   \nonumber \\
	& \times e^{-\frac{1-T}{2} (1+\frac{\varepsilon}{2})|\xi_1- \xi_2 + \frac{\sqrt{2}}{\tau} \xi_\mathrm{N}|^2}    \nonumber \\
	& \times e^{-\frac{1-\mu}{2}|\xi_1 + \xi_2 -\frac{\sqrt{2}}{g} \xi_3|^2 } \nonumber \\
	& \times e^{-\frac{\mu}{2}|\xi_1 + \xi_2 + \sqrt{2} {g} \xi_3|^2 } \nonumber \\
	& \times \Big( 1-\frac{\mu}{2}   |\xi_1 + \xi_2 + \sqrt{2} {g} \xi_3|^2 \Big),
\end{align}
where $g=\sqrt{(1-\mu)/\mu}$, $\tau=\sqrt{(1-T)/T}$, and  $ \text{Im}[\xi]$ being the imaginary part of complex number $\xi$.
Using \eqref{rec-state}, the joint state of the output modes is then given by
\begin{align}
	\hat{\rho}_{\mathcal{B}} = & \int \frac{d^2\xi_1}{\pi}  \int \frac{d^2\xi_2}{\pi} \int \frac{d^2\xi_3}{\pi} \int \frac{d^2\xi_{\rm N}}{\pi}  \chi_{\mathrm{A}}^{\mathrm{out}} (\xi_1, \xi_2, \xi_3, \xi_\mathrm{N}) \nonumber \\ & \hat{D}_{\mathrm{N}}(\hat{b}_1, \xi_1)  \hat{D}_{\mathrm{N}}(\hat{b}_2, \xi_2)  \hat{D}_{\mathrm{N}}(\hat{b}_3, \xi_3) \hat{D}_{\mathrm{N}}(\hat{b}_\mathrm{N}, \xi_\mathrm{N}).
\end{align}
We can next trace out mode $\hat{b}_{\rm N}$ to obtain the joint state of $[\hat{b}_1 ~\hat{b}_2 ~\hat{b}_3 ]$, which is 
\begin{align}
	\hat{\rho}_{\mathrm{out}} = & \int \frac{d^2\xi_1}{\pi}  \int \frac{d^2\xi_2}{\pi} \int \frac{d^2\xi_3}{\pi} \chi_{\mathrm{A}}^{\mathrm{out}} (\xi_1, \xi_2, \xi_3, 0) \nonumber \\ & \hat{D}_{\mathrm{N}}(\hat{b}_1, \xi_1)  \hat{D}_{\mathrm{N}}(\hat{b}_2, \xi_2)  \hat{D}_{\mathrm{N}}(\hat{b}_3, \xi_3),
\end{align}
where 
\begin{align}
	\chi_{\mathrm{A}}^{\mathrm{out}} (\xi_1, \xi_2, \xi_3,0) = &
	e^{-F_1|\xi_1- \xi_2|^2} e^{\sqrt{2T} i  \text{Im}[\bar{\alpha} (\xi_1- \xi_2) ] }   \nonumber \\
	& \times  e^{-\frac{\mu}{2}|\xi_1+\xi_2+\sqrt{2}{g}\xi_3|^2} e^{-\frac{1-\mu}{2}|\xi_1+\xi_2-\frac{\sqrt{2}}{g}\xi_3|^2} \nonumber \\
	& \times  \big(1 - \frac{\mu}{2}|\xi_1+\xi_2+\sqrt{2}{g}\xi_3|^2\big) ,
\end{align}
with $F_1=\frac{1}{2}+ \frac{1}{4} (1-T)\varepsilon $. Note that $\varepsilon_{\rm rec}=(1-T)\varepsilon$ is the amount of excess noise at the receiver side; thus, $F_1=\frac{1}{2}+ \frac{1}{4} T\varepsilon_{\rm tm}$, where $\varepsilon_{\rm tm} = \varepsilon_{\rm rec}/T$ is the amount of excess noise at the transmitter.

\subsection{Post-selected state}
\label{subsec:post-sel}
Following \cite{Ralph_Lund_QSNLA}, we consider a QS to be  successful if only one detector in Fig.~\ref{fig:PandM} clicks. In order to model such measurements we use the following non-resolving measurement operator 
\begin{align}
	\label{meas-op}
	\hat{M} = (\mathbbm{1} -|0\rangle_1\langle 0|)\otimes |0\rangle_2\langle 0|,
\end{align}
which corresponds to the case where detector D1 clicks while D2 does not. The post-selected state, $\hat{\rho}_{\mathrm{out}}^{\mathrm{PS}}$, is then given by \cite{Nielsen_Chuang}:
\begin{align}
	\label{PM-out}
	\hat{\rho}_{\mathrm{out}}^{\mathrm{PS}} = & \frac{\mathrm{tr}_{\hat{b}_1\hat{b}_2} (\hat{\rho}_{\mathrm{out}} \hat{M} )}{\mathrm{tr} ( \hat{\rho}_{\mathrm{out}} \hat{M})  }   \nonumber \\
	= &  \frac{1}{P^{\mathrm{PS}}} \int \frac{d^2\xi_1}{\pi}  \int \frac{d^2\xi_2}{\pi} \int \frac{d^2\xi_3}{\pi}  \chi_{\mathrm{A}}^{\mathrm{out}} (\xi_1, \xi_2, \xi_3,0)  \nonumber \\
	& \times (\pi\delta^2(\xi_1) -1) \hat{D}_{\mathrm{N}}(\hat{b}_3, \xi_3) ,
\end{align}
where $\delta^2(\xi)=\delta(\xi_r)  \delta(\xi_i)$ and $P^{\mathrm{PS}}= \mathrm{tr} (\hat{M}  \hat{\rho}_{\mathrm{out}} )$ is the corresponding (success) probability of the measurement $\hat{M}$, which will be calculated in Sec.~\ref{subsec:succ_prob}.

Because the truncated post-measurement state lives in the qubit subspace spanned by number states $\{ |0\rangle_{\hat{b}_3},|1\rangle_{\hat{b}_3}\}$, the output state has the form 
\begin{align}
	\label{PS-state}
	\hat{\rho}_{\mathrm{out}}^{\mathrm{PS}} (\alpha) = & \rho_{00}  (\alpha) |0\rangle_{\hat{b}_3}\langle 0| + \rho_{01}  (\alpha) |0\rangle_{\hat{b}_3}\langle 1| \nonumber \\
	& + \rho_{10} (\alpha) |1\rangle_{\hat{b}_3}\langle 0| + \rho_{11}  (\alpha) |1\rangle_{\hat{b}_3}\langle 1|  ,
\end{align}
where $\rho_{jk} (\alpha) = {}_{\hat{b}_3}{\langle j }|  \hat{\rho}_{\mathrm{out}}^{\mathrm{PS}}  (\alpha) |k \rangle_{\hat{b}_3}$, for $j,k=0,1$. 
We then obtain
\begin{align}
	\label{state_coeff}
	\begin{cases}
		\rho_{00}(\alpha)= \frac{2[2F_1(2F_1+1)+T|\alpha|^2]}{(g^2+1)(2F_1+1)^3} e^{-T\frac{|\alpha|^2}{2F_1+1}} / P^{\mathrm{PS}} (\alpha)  \\
		\rho_{01}(\alpha)= \frac{-2g\sqrt{T}\alpha}{(g^2+1)(2F_1+1)^2} e^{-T\frac{|\alpha|^2}{2F_1+1}} /P^{\mathrm{PS}}(\alpha) = \rho_{10}^\ast(\alpha)  \\
		\rho_{11}(\alpha)=   \frac{2g^2}{g^2+1} \Big( \frac{e^{-T\frac{|\alpha|^2}{2F_1+1}}}{2F_1+1}  - \frac{e^{-T\frac{|\alpha|^2}{2F_1}}}{4F_1}   \Big)  /P^{\mathrm{PS}}(\alpha) .            
	\end{cases}
\end{align}

We remark that in the case that only detector D2 clicks, the QS is still considered successful. After working out the post-selected output state, we find that the result has the same form as in \eqref{PS-state}, but we only need to replace $\alpha$ with $-\alpha$ in \eqref{state_coeff}. In practice, in a QKD setup, Bob can negate its measurement results whenever this happens. One can also use a unitary operation to correct the output state so that we always end up with \eqref{PS-state} as the post-selected state.

We note that the post-measurement state is Hermitian and positive-semidefinite, as expected. In addition, in the limit of $|g\alpha|\ll 1$,  we can verify that the post-selected state of the single QS approaches the weak coherent state $| g\alpha\rangle $.

\subsection{Probability of success}
\label{subsec:succ_prob}
The probability of success for measurement $\hat M$ and input $|\alpha\rangle$ is given by
\begin{align}
	P^{\mathrm{PS}}(\alpha)= & \mathrm{tr} (\hat{\rho}_{\mathrm{out}} \hat{M}  ) \nonumber \\
	= & \int \frac{d^2\xi_1}{\pi} \int \frac{d^2\xi_2}{\pi}  \chi_{\mathrm{A}}^{\mathrm{out}} (\xi_1, \xi_2, 0,0) (\pi\delta^2(\xi_1) -1) .
\end{align}
By substituting \eqref{A-func-out} into the above expression, we obtain
\begin{align}
	\label{succ-prob}
	P_{\mathrm{succ}}(\alpha)= & 2P^{\mathrm{PS}}(\alpha) \nonumber \\
	= & \frac{4\big(g^2(2F_1+1)^2 +2F_1(2F_1+1)+T|\alpha|^2\big)}{(g^2+1)(2F_1+1)^3}   \nonumber \\
	& \times e^{-T\frac{|\alpha|^2}{2F_1+1}}- \frac{g^2 }{(g^2+1)F_1} e^{-T\frac{|\alpha|^2}{2F_1}},
\end{align}
where $P_{\mathrm{succ}}(\alpha)$ is the total probability of success for the QS module, i.e., when either of D1 or D2 detector clicks. As expected, $P_{\mathrm{succ}}(\alpha)$ approaches, to first-order approximation, to $P_{\mathrm{succ}}^{\mathrm{RL}}(\alpha)= \mu + (1-\mu)|\alpha|^2 = (1+|g\alpha|^2)/(1+g^2)$, when $|\alpha|\ll 1$, at $\varepsilon=0$ and $T=1$.

\begin{figure}[t]
	\centering
	\includegraphics[scale=0.45]{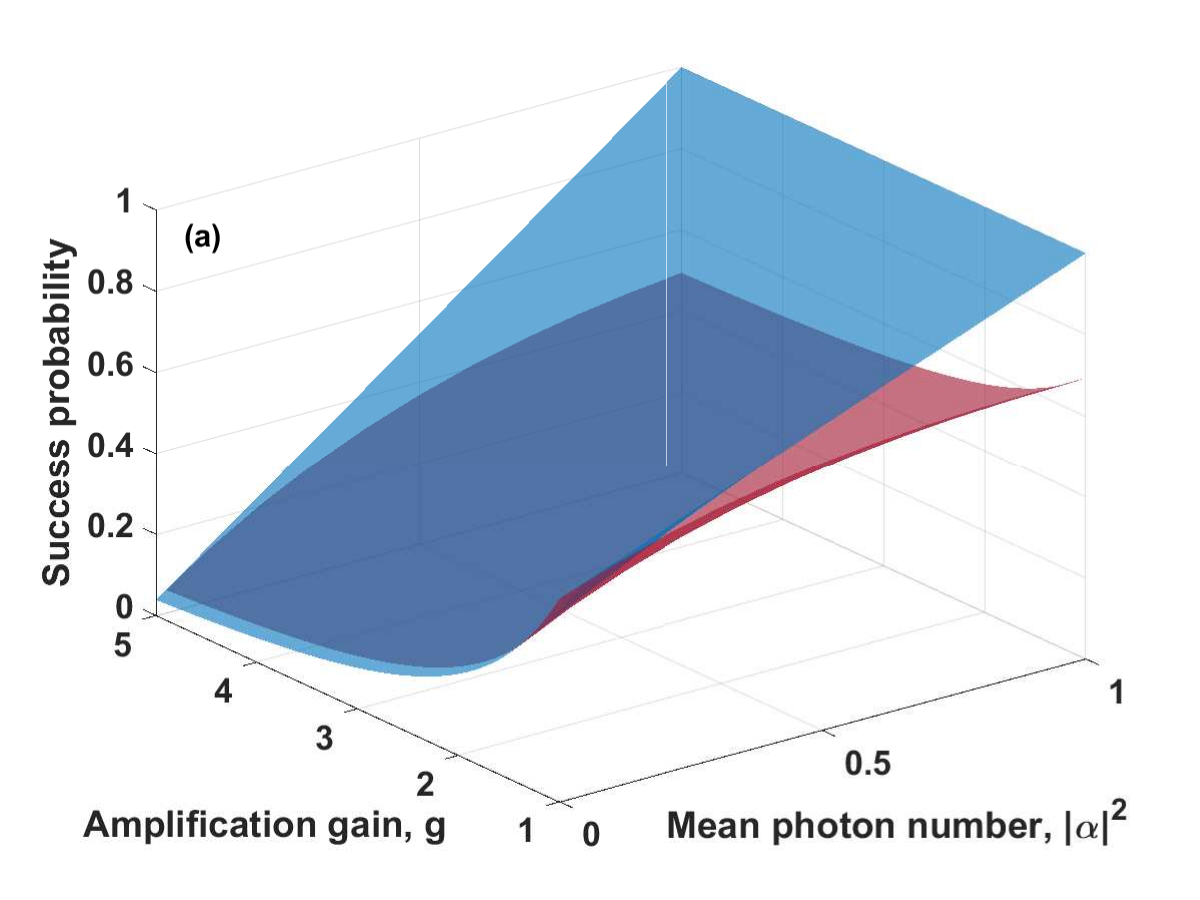} \hspace{0.8cm} \includegraphics[scale=0.45]{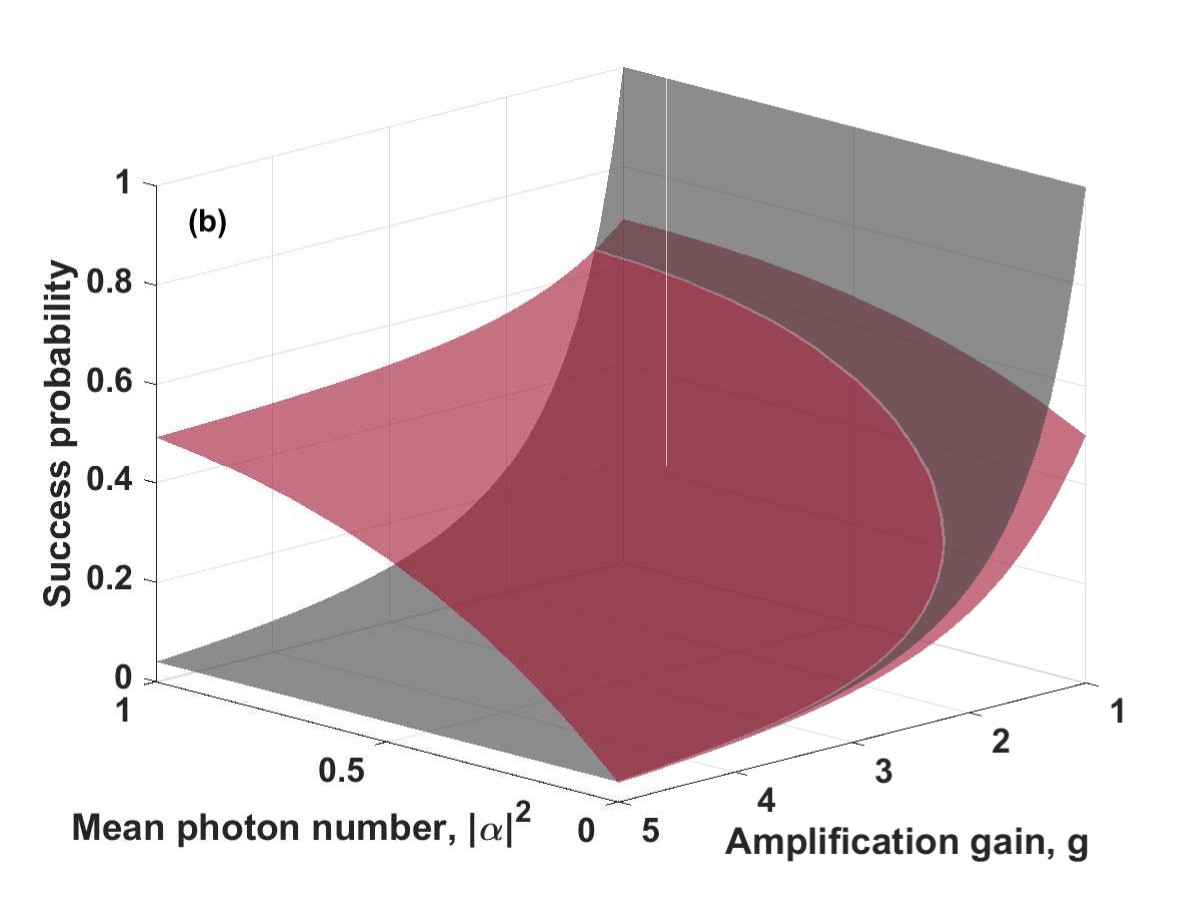}
	\caption{(a) The exact success probability of a single QS (lower red), $P_{\mathrm{succ}}$, and that based on approximations in  \cite{Ralph_Lund_QSNLA} (upper blue), $P_{\mathrm{succ}}^{\mathrm{RL}}$. (b) The exact success probability of a single QS (red), $P_{\mathrm{succ}}$, and that of an ideal NLA (grey), upper bounded by $1/g^2$, versus average photon number and amplification gain.  In all cases,  $\varepsilon=0$ and $T=1$.}
	\label{fig:prob-succ}
\end{figure}

This approximation is, however, invalid even when we slightly deviate from the condition on $|\alpha|$, as can be seen in Fig.~\ref{fig:prob-succ}(a). Here, we have plotted the exact probability of success, $P_{\mathrm{succ}}(\alpha)$, versus $|\alpha|^2$ and $g$, and compared it with the asymptotic value obtained by Ralph and Lund, $P_{\mathrm{succ}}^{\mathrm{RL}}(\alpha)$. It can be seen that the exact probability of success is always lower than the asymptotic value, and the difference is visible at all values of $g$. The success probability also increases with the decrease in $g$. For $|\alpha|\ll 1$, the success probability approaches its maximum possible value of $1/g^2$ \cite{Pandey_Qlimits_Amp}. But, again, as can be seen in Fig.~\ref{fig:prob-succ}(b), we quickly deviate from this ideal regime when $|\alpha|$ increases. This indicates that we cannot operate at maximum possible success probability for all possible inputs, as assumed in  \cite{Blandino_CVQKD_idealNLA}, if we use a QS as an NLA. 

%------------------------------------------------------------
%------------------------------------------------------------ 

In Fig.~\ref{fig:prob-succ}(b), the maximum possible success probability, $1/g^2$, divides the plot into two regions. There is a region in which the success probability is above the maximum possible for an NLA. This implies that the QS operation should be very noisy in this region, hence breaking the assumption on the noise-free operation of the NLA. If we want to work in the region that $P_{\mathrm{succ}}(\alpha) < 1/g^2$, we will then have to deal with limitations on the maximum gain that we can choose for the range of input states we may expect. This indicates a trade-off between the amount of noise that the QS may add to the signal versus its gain and success probability. We will later address this issue, in the context of CV\:QKD, in our numerical results when we optimize the secret key generation rate over the system parameters.

\subsection{Non-Gaussian behavior of the QS}
\label{subsec:non-Gaussianity}
Before calculating the secret key generation rate of a QS-equipped CV\:QKD system, it is necessary to better understand the nature of a quantum channel that includes a QS module. This is important because the majority of results on the secret key rate of CV\:QKD systems rely on Gaussian characteristics of the channel \cite{Garcia-Patron_OptGaussAttacks,Lodewyck_25kmCVQKD}. This is not, however, the case for a QS module as we see in this section. 

In order to examine the non-Gaussian behavior of the QS output, let us focus on the distribution of homodyne measurement results on quadrature $\hat x_B$. Let us also consider an input coherent state $|\alpha\rangle $, with $\alpha=x_A+ip_A$ as distributed by \eqref{Eq:InputDist}, at the QS port $\hat{a}_1$.
That results in a thermal state with variance $V_A$ and given by $\int d^2\alpha  \frac{e^{-\frac{|\alpha|^2}{V_A/2}}}{\pi V_A/2} |\alpha\rangle_{\hat{a}_1}\langle\alpha|$. 
After performing similar calculations, the post-selected state will be given by
\begin{align}
	\hat{\sigma}_{\mathrm{out}}^{\mathrm{PS}}(V_A) = & \sigma_{00}(V_A) |0\rangle_{\hat{b}_3}\langle 0| + \sigma_{11}(V_A)  |1\rangle_{\hat{b}_3}\langle 1| ,
\end{align}
where 
\begin{align}
	\label{state_coeff_th}
	\begin{cases}
		\sigma_{00}(V_A)= \frac{8F_2}{(g^2+1)(2F_2+1)^2P_{\mathrm{succ}} (V_A)}   \\
		\sigma_{11}(V_A)= \frac{4g^2}{(g^2+1)P_{\mathrm{succ}} (V_A)}  \big( \frac{1}{2F_2+1} - \frac{1}{4F_2} \big)            
	\end{cases}
\end{align}
with success probability 
\begin{align}
	P_{\rm succ}(V_A)= \frac{4}{(g^2+1)}\Big( \frac{g^2(2F_2+1)+2F_2}{(2F_2+1)^2}    - \frac{g^2}{4F_2} \Big)
\end{align}
and $F_2=\frac{1}{2} + \frac{1}{4}T(V_A+\varepsilon_{\rm tm}) $. 

%------------------------------------------------------------
\begin{figure}[t]
	\includegraphics[scale=0.52]{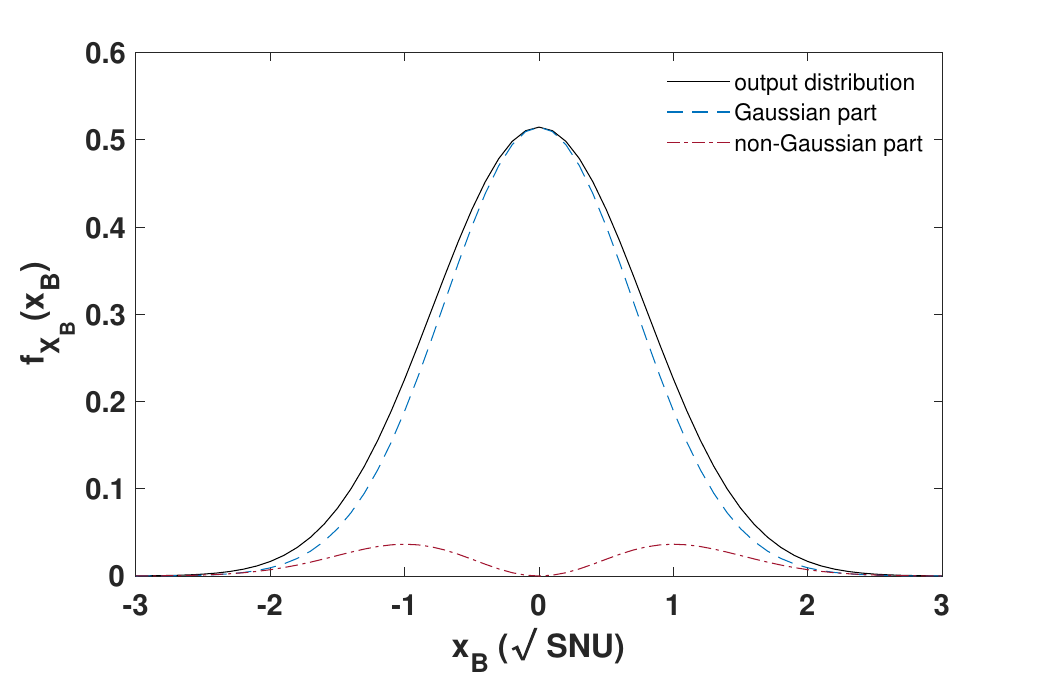}
	\caption{The output distribution at the receiver side (solid-black), which comprises Gaussian (dashed blue) and non-Gaussian (dot-dashed red) parts. Here, $g=2$, $V_A= 0.05$, $\varepsilon=0$, and $T=1$.}
	\label{fig:non-Gauss}
\end{figure}
%------------------------------------------------------------ 

The probability distribution for obtaining a real number $x_B$ after measuring $\hat{x}_B$, conditional on the success of the QS, is then given by
\begin{align}
	\label{output_dist}
	f_{X_B}(x_B) &= \mathrm{tr} ( \hat{\sigma}_{\mathrm{out}}^{\mathrm{PS}}(V_A) |x_B \rangle \langle x_B | ) \nonumber \\
	&=  \big( \sigma_{00}(V_A) + 2\sigma_{11}(V_A) x_B^2 \big) \frac{e^{-x_B^2} }{\sqrt{\pi}},
\end{align}
where $\hat x_B |x_B\rangle = x_B |x_B\rangle$.

The expression for $f_{X_B}(x_B)$ will then have two components: one is a Gaussian term in $x_B$ proportional to $\sigma_{00}(V_A)$, and the other is a non-Gaussian term proportional to $\sigma_{11}(V_A)$. Fig.~\ref{fig:non-Gauss} shows the contribution of each of these components in making $f_{X_B}(x_B)$ at $g=2$, $V_A = 0.05$, $\varepsilon=0$, and $T=1$. We notice that even for such a small modulation variance, which corresponds mostly to small values of $|\alpha|$, the non-Gaussian term is quite distinct. Higher amplification gains could even result in more deviation from a Gaussian state. This non-Gaussian behavior would have ramifications on the key rate analysis of a QS-based system as we see next.

\section{Secret key rate analysis}
\label{sec:secret_key_analysis}
In this section, we use the results in Sec.~\ref{sec:quantum_scissor} to determine the secret key rate of the GG02 protocol when Bob uses a single QS before his homodyne measurement. We find the secret key rate in a nominal operation condition when no eavesdropper is present. We, however, assume a thermal loss channel with transmissivity $T$, modeled by a beam splitter, and an excess noise $\varepsilon$. This can effectively be thought as having an eavesdropper who attempts a Gaussian attack \cite{Pirandola:PRL_GaussAtt2008}. That is, we assume that Eve employs an entangling cloner by coupling one component of a TMSV state with Alice's signal, while retaining the other part for her future measurements. If one traces out the latter part of the TMSV state that Eve would keep for herself, the state on the other part would be a thermal state. The effective impact of Eve’s attack on the channel will then be equivalent to coupling Alice’s signal with a thermal state, which is the same as using a thermal-loss channel for analysing the secret key rate, as we have pursued in this work. It is important to note that such an attack may not be the optimal one for our non-Gaussian channel, but considering how close the output distribution in Fig.~\ref{fig:non-Gauss} is to a Gaussian distribution, the results obtained for this particular channel should not be far away from that obtained in an optimal attack \cite{Malaney_Sat}. The secret key rate of CV\:QKD protocols in the asymptotic limit of infinitely many signals is given by 
\begin{align}
	\label{rate}
	K= \beta I_{\rm AB} - \chi_{\rm BE},
\end{align}
where $\beta$, $I_{\rm AB}$, $\chi_{\rm BE}$ are, respectively, the reconciliation efficiency, the mutual information between Alice and Bob, and eavesdropper's accessible information when reverse reconciliation is used. 

In our proposed setup, since the QS operation is non-deterministic, the whole key rate formula should be multiplied by the \emph{average} success probability of the QS, $\overline{P}_{\mathrm{succ}}$, where the averaging is performed over all possible inputs. Therefore, the secret key rate reads 
\begin{align}
	\label{rate-NLA}
	K_{\mathrm{QS}} \geq \overline{P}_{\mathrm{succ}} (\beta I_{\rm AB}^\star - \chi_{\rm BE}^\star),
\end{align}
where `$\star$' indicates that the mutual and Holevo information terms are calculated for the post-selected data when the QS is successful. The measurement results corresponding to unsuccessful QS events will be discarded at the sifting stage.   

The fact that we only use the post-selected data for key extraction implies that we have to account for the non-Gaussianity of the QS output states. Unfortunately, the non-Gaussian behavior of the QS makes conventional methods for key rate calculation inapplicable. In order to take the non-Gaussian effects into account, we calculate the exact mutual information by directly using the conditional distribution of the QS output. 
Ideally one could also look for the exact calculation of the Holevo information term as well. But, this turns out to be extremely cumbersome. Instead, in this paper, we find an upper bound for the Holevo information term  by finding the covariance matrix (CM) of the output state from the total channel and then calculate the Holevo information for a Gaussian state with the same CM. 
The reason is that Gaussian collective attacks are proven to be optimal in the sense that they maximize the Holevo quantity \cite{Garcia-Patron_OptGaussAttacks} of fixed CM for the output shared state. 
Given the generality of the results in \cite{Garcia-Patron_OptGaussAttacks}, in a real experiment, once we obtain the CM terms from the measurement results, we can use the same methodology to obtain a lower bound on the key rate.

In the following, we provide more detail on how each of the terms in \eqref{rate-NLA} can be calculated.

\subsection{Mutual information}
The mutual information between two random variables $X_A$ and $X_B$, corresponding to post-selected data on Alice's and Bob's sides, is the difference between the entropy function $H(X_B)$ and the conditional entropy $H(X_B|X_A)$ \cite{Cover}:
\begin{align}
	\label{mutinf_def}
	I_{\rm AB}^\star= H(X_B)-H(X_B|X_A),
\end{align}
where
\begin{align}
	\label{Ent-XB}
	H(X_B)=- \int dx_B ~ f_{X_B}(x_B)  \log_2 f_{X_B}(x_B) ,
\end{align}
and
\begin{align}
	\label{Ent-XB-cond}
	H(X_B|X_A)=& -  \int \int dx_A dx_B  f(x_A,x_B) \log_2 f_{X_B}(x_B|x_A),
\end{align}
with $ f(x_A,x_B)= f_{X_A}(x_A) f_{X_B}(x_B|x_A) $ being the joint probability density function. 

Here, $f_{X_B}(x_B)$ can be obtained by using \eqref{output_dist}, while the conditional output distribution $f_{X_B}(x_B|x_A)$ can be obtained as follows:
\begin{align}
	f_{X_B}(x_B|x_A) &= \mathrm{tr} ( \hat{\omega}_{\mathrm{out}}^{\mathrm{PS}}(x_A) |x_B \rangle \langle x_B | ),
\end{align}
where the conditional output state $\hat{\omega}_{\mathrm{out}}^{\mathrm{PS}}(x_A) $ is calculated in Appendix~\ref{app:output_dist_cond}.
In our work, we numerically carry out the above integrals for a given set of parameters.

\subsection{Holevo information} 
\begin{figure}[t]
	\centering
	\includegraphics[scale=0.75]{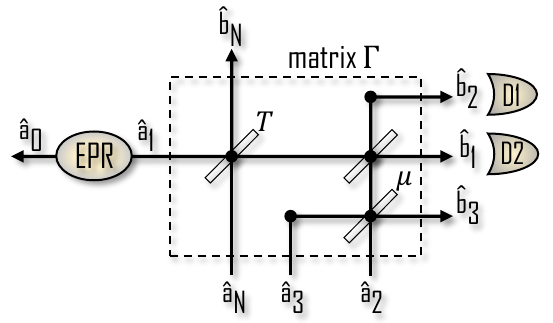}
	\caption{QS-amplified EB CV\:QKD scheme. The quantum channel and the QS are considered as a combined system with input modes $\hat{a}_1 - \hat{a}_3$ and $\hat{a}_\mathrm{N}$ and output modes $\hat{b}_1 - \hat{b}_3$ and $\hat{b}_\mathrm{N}$. The transformation matrix of the system is given by \eqref{matrix-TQS}.}
	\label{fig:EPRQS}
\end{figure}

In order to calculate the Holevo information term, $\chi_{\rm BE}^\star$, we use the EB description of the protocol, where one part of an EPR state travels through the quantum channel and amplified by a QS, while the other is measured by Alice; see Fig.~\ref{fig:EPRQS}. In order to upper bound $\chi_{\rm BE}^\star$, what we need is then the CM of Alice-Bob bipartite state. We will then first derive the exact post-selected joint state, from which the CM parameters can be obtained. We use a similar approach to Sec.~\ref{sec:quantum_scissor} in using characteristic functions to find a relationship between Alice and Bob states when the QS is successful. As shown in Fig.~\ref{fig:EPRQS}, we also account for the effect of the quantum channel loss and excess noise in our calculations.

By using \eqref{antinorm-func} and the transformation matrix $\Gamma$, we can now write the full output antinormally-ordered characteristic function, including $\hat{a}_0$ mode, in terms of the input one by $\chi_{\mathrm{A}}^{\mathrm{out}}(\xi_0, \xi_1,\xi_2, \xi_3 ,\xi_\mathrm{N}) = \chi_{\mathrm{A}}^{\mathrm{in}}(\lambda_0, \lambda_1,\lambda_2, \lambda_3 ,\lambda_\mathrm{N})$, where 
\begin{align}
	[\xi_0~ \xi_1~\xi_2~ \xi_3 ~\xi_\mathrm{N} ] =
	\left(\begin{array}{cccc}
		1 & 0 \\
		0 &  \Gamma   
	\end{array}\right)
	[\lambda_0~ \lambda_1~\lambda_2~ \lambda_3~\lambda_\mathrm{N}] , \nonumber
\end{align}
with $\chi_{\mathrm{A}}^{\mathrm{in}}(\lambda_0, \lambda_1,\lambda_2, \lambda_3 ,\lambda_\mathrm{N}) = \chi_{\mathrm{A}}^{\mathrm{EPR}}(\lambda_0,\lambda_1) \times \chi_{\mathrm{A}}^{\mathrm{in}}(\lambda_2,\lambda_3,\lambda_\mathrm{N})$,
where $	\chi_{\mathrm{A}}^{\mathrm{EPR}} (\lambda_0,\lambda_1) = \exp \{-\delta^2 (|\lambda_0|^2+|\lambda_1|^2) -2 \text{Re} (\delta \gamma  \lambda_0^\ast \lambda_1^\ast ) \} $ is the antinormally-ordered characteristic function of the EPR state with parameters $\delta$ and $\gamma=\sqrt{\delta^2-1}$, and $ \text{Re}[\xi]$ being the real part of the complex number $\xi$.
The term $\chi_{\mathrm{A}}^{\mathrm{in}}(\lambda_2,\lambda_3,\lambda_\mathrm{N})$ is calculated for input state $|1\rangle_{\hat{a}_2}\langle 1| \otimes |0\rangle_{\hat{a}_3} \langle 0 | \otimes \int d^2\beta f_{\varepsilon}(\beta) |\beta \rangle_{\hat{a}_\mathrm{N}} \langle \beta |$. 

Putting all this together, we then find the pre-measurement antinormally-ordered characteristic function for modes $\hat{a}_0$, $\hat{b}_1$, $\hat{b}_2$, $\hat{b}_3$, and $\hat{b}_\mathrm{N}$ as follows:
\begin{align}
	\label{EPRout-anti-nor}
	\chi_{\mathrm{A}}^{\mathrm{out}} (\xi_0, \xi_1,\xi_2,\xi_3 ,\xi_\mathrm{N})= &  e^{-\delta^2 |\xi_0|^2}  e^{-\kappa  \text{Re}\big( \xi_0^\ast (\xi_1^\ast - \xi_2^\ast )  \big) }  \nonumber \\
	&\times  e^{-\frac{\delta^2 T}{2}|\xi_1- \xi_2 -\sqrt{2}\tau \xi_\mathrm{N}|^2 }    \nonumber \\
	&\times 	e^{-\frac{1-T}{2} (1+\frac{\varepsilon}{2})|\xi_1- \xi_2 + \frac{\sqrt{2}}{\tau} \xi_\mathrm{N}|^2}    \nonumber \\
	& \times e^{-\frac{1-\mu}{2}|\xi_1 + \xi_2 -\frac{\sqrt{2}}{g} \xi_3|^2 } \nonumber \\
	&\times  e^{-\frac{\mu}{2}|\xi_1 + \xi_2 + \sqrt{2} {g} \xi_3|^2 } \nonumber \\
	& \times \Big( 1-\frac{\mu}{2}   |\xi_1 + \xi_2 + \sqrt{2} {g} \xi_3|^2 \Big),
\end{align}
where  $\kappa=2\delta\gamma \sqrt{T/2}$.

In the EB scheme, we find the corresponding parameter $\delta$ in our EPR state, which gives the same output statistics for the signal that goes to Bob, when Alice does a heterodyne measurement on her state. It then turns out that to get an identical output state we should satisfy $\delta= \sqrt{(V+1)/2}$, where $V = V_A + 1 $. 

Having obtained the output characteristic function, we can find the corresponding output density matrix using \eqref{rec-state}. Then, by tracing out the output mode $\hat{b}_\mathrm{N}$ and also performing photon-detection measurements on modes $\hat{b}_1$ and $\hat{b}_2$---by introducing the same measurement operator as in \eqref{meas-op}---we find the resultant joint state of $\hat{a}_0$ and $\hat{b}_3$ modes in the case of having a successful event. 

Appendix \ref{app:CM_elements} provides the detailed calculations of the post-measurement density matrix, and the corresponding CM parameters. It turns out that the CM of the shared bipartite state between Alice and Bob has the form 
\begin{align}
	\label{CM-EPR}
	\gamma_{\rm AB}=
	\left(\begin{array}{cc}
		a \mathbbm{1} & c \sigma_{\mathrm{z}} \\
		c \sigma_{\mathrm{z}} & b \mathbbm{1}
	\end{array}\right),
\end{align}
where $\mathbbm{1}=\text{diag}(1,1)$ and $\sigma_{\mathrm{z}}=\text{diag}(1,-1)$ with 
\begin{align}
\label{abc}
a= & \Big(\frac{8[ \gamma^2 T + \big(2F_3+1- \gamma^2  T \big) \big(g^2 (2F_3+1)+2F_3 \big) ] }{(2F_3+1)^3}  \nonumber \\
& - \frac{ g^2 (2F_3-\gamma^2 T)}{F_3^2} \Big)  \frac{\delta^2}{(g^2+1)\overline P_{\rm succ}}  -1 , \nonumber \\
b= &\frac{4}{(g^2+1) \overline P_{\rm succ} }  \Big( \frac{4[g^2 (2F_3+1)+F_3] }{(2F_3+1)^2}   - \frac{g^2}{F}  \Big) -1 , \nonumber \\
c= & \frac{8\delta \gamma}{(g^2+1)\overline P_{\rm succ} (2F_3+1)^2 }  g\sqrt{T} ,
\end{align}
$F_3=\frac{1}{2}+\frac{1}{4}T (2(\delta^2-1) + \varepsilon_{\rm tm} )$ and $\overline P_{\rm succ} = \frac{1}{g^2+1} \big( 4[(2F_3+1)g^2+2F_3]/(2F_3+1)^2 - g^2/F_3 \big)$.

It is interesting to make the following observation. 
If the EPR state is assumed totally uncorrelated, which happens when its squeezing parameter goes to zero, both parts of the state are left with vacuum states. Thus, if the QS is successful, the output state of mode $\hat{b}_3$ should be a vacuum state as well. This means that the CM of the end-to-end state is the identity \cite{Weedbrook_GaussQIRev_2012}. We verify that in the case of having a totally uncorrelated EPR state, corresponding to $\delta=1$ and $\gamma=0$, the expressions above will indeed result in the identity matrix; that is, we obtain $a=b=1$ and $c=0$. 

In addition, as a result of the statistical equivalence between EB and PM schemes, where $\delta=\sqrt{(V+1)/2}$, we conclude that $F_3=F_2$.
Now that the CM is known, we can upper bound the Holevo information by using \eqref{app:Eq-Holevo}. 

%------------------------------------------------------------
%------------------------------------------------------------

\section{Numerical results}
\label{sec:numerical_results}
In this section, we present numerical simulations of the secret key rate of the QS-amplified GG02 protocol and compare it with that of the conventional one. We find the maximum value for the lower bound in \eqref{rate-NLA} by optimizing, at each distance, 
the modulation variance, $V_A$, or, equivalently, the parameter $\delta$ in the EB scenario, as well as the QS parameter, $\mu$, which specifies the QS amplification gain. We also account for the excess noise, as discussed in previous sections. We assume that the quantum channel between the sender and receiver is an optical fiber with loss factor $\upalpha$, whose transmittance is given by $T=10^{-\upalpha L/10}$, where $L$ is the channel length and the loss factor is $\upalpha=0.2$ dB/km corresponding to standard optical fibers. Also, we assume $\beta=1$ and that ideal homodyne detection, with no electronic noise, is performed at the receiver. 

\begin{figure}[b]
	\includegraphics[width=1.08 \linewidth]{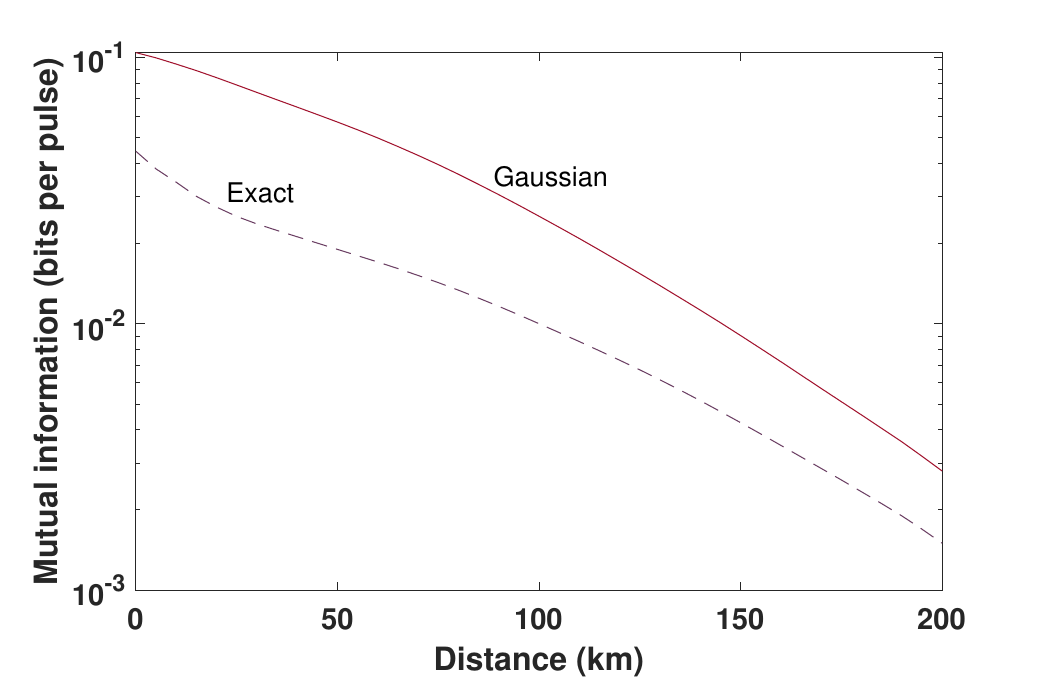}
	\caption{The exact mutual information function (dashed) as compared to its Gaussian approximation (solid) versus distance at $\varepsilon = 0$. All other parameters have been optimized.}
	\label{fig:mutual_info}
\end{figure}

We first highlight the importance of accounting for the non-Gaussian behavior of the QS by comparing the difference between the exact value of the mutual information function $I_{\rm AB}^\star$, given by \eqref{mutinf_def}, and that obtained by Gaussian approximation, $I_{\rm AB}^{\rm G}$, in \eqref{Eq:IABG}. Fig.~\ref{fig:mutual_info} shows both curves, versus distance, at no excess noise.
It is clear that the Gaussian approximation would have overestimated the mutual information between Alice and Bob at all distances considered, and that could have resulted in wrong bounds for the key rate of QS-based systems.

%------------------------------------------------------------
%----------------------------------------------------------- 

\begin{figure}[t]
	\includegraphics[width=1.08\linewidth]{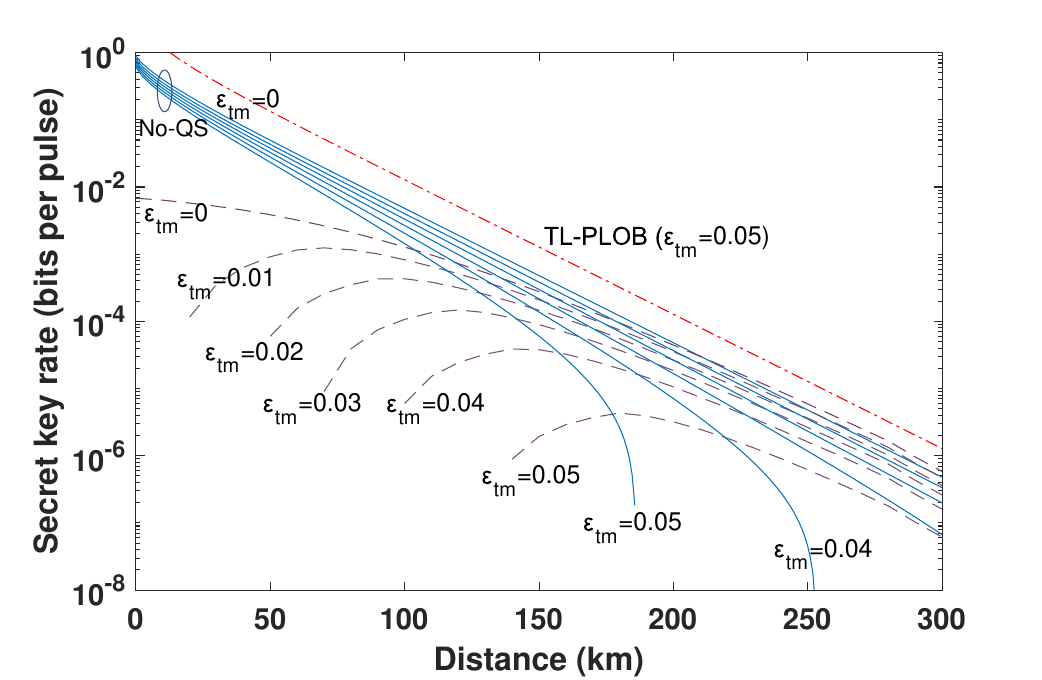}
	\caption{The optimized secret key rate for the QS-amplified CV\:QKD protocol versus distance, as compared to the rate of conventional GG02, and the upper bound for a repeaterless thermal-loss channel (TL-PLOB) at a mean thermal photon number of $\varepsilon_{\rm rec}/(2(1-T))$. The solid lines represent the no-QS case with top curve at $\varepsilon_{\rm tm} = 0$, and the bottom one at $\varepsilon_{\rm tm} = 0.05$, and the middle curves covering $\varepsilon_{\rm tm} = 0.01-0.04$}. 
	\label{fig:rates}
\end{figure}

Figure~\ref{fig:rates} shows the optimized secret key rates of both conventional (solid lines) and the QS-assisted (dashed lines) GG02 protocol versus distance, as well as that of the PLOB bound for a repeaterless thermal-loss channel (labelled TL-PLOB) \cite{Pirandola_PLOB17}. This is the bound given in (23) of  \cite{Pirandola_PLOB17} at an equivalent mean thermal photon number, $\bar n =  \varepsilon_{\rm tm} T/(2(1-T))$, to our receiver excess noise (here at $\varepsilon_{\rm tm} = 0.05$). There are several interesting observations that can be made in this figure. First, we note that for all considered cases, there exists a cross-over distance at which the QS-assisted curves surpass their corresponding no-QS curves.
At $\varepsilon_{\rm tm} = 0$, this happens at around 200~km. By increasing $\varepsilon_{\rm tm}$, the cross-over distance would drop and reaches around 150~km at $\varepsilon_{\rm tm} = 0.05$. This proves the key objective of our work that, by using realistic NLAs, there would be certain regimes where NLA-based systems improve the performance and the distance at which secure keys can be exchanged.  

%------------------------------------------------------------
%------------------------------------------------------------ 

It can be seen, in Fig.~\ref{fig:rates}, that QS-equipped receivers may not support high key rates at short distances. In fact, except for the case of $\varepsilon_{\rm tm} = 0$, we may not be able to exchange any secret keys at very short distances for the QS-based system. Even for the no excess noise case, there are over two orders of magnitude difference between the no-QS and QS-based curves at $L=0$. This is attributed to multiple factors. First, the trade-off between the choice of modulation variance and noise level in the system, would require us to use very small values of $V_A$ at short distances, otherwise the QS will not operate at its low-noise regime. For instance, at $L=0$, the optimum value of $V_A$ for the QS-based system is 0.04. A no-QS system with such a low value of $V_A$ also offers a low key rate of $2.83 \times 10^{-2}$, which is comparable to what we obtain for the QS-based system. Another factor is the success probability that at $L=0$ is around 0.5, and it almost linearly goes down to around 0.15 at 200~km. One last factor is also the fact that the QS is not entirely noise free. The additional noise by the QS would further decrese the rate at $L=0$. In addition to this, if we have nonzero values of excess noise, a combination of the above effects plus the external noise drive the key rate to zero at very short distances. This is by itself is not a practical dilemma, as, for a given channel length, one, in advance, can figure out whether to use a QS or not. But, this can affect the applicability of QS modules in a CV quantum repeater system.

Another observation in Fig.~\ref{fig:rates} is that, at long distances, the key rate for QS-based systems follows a parallel trend to that of the TL-PLOB curve. For instance, at $\varepsilon_{\rm tm} = 0.05$, the key rate remains roughly one order of magnitude below the PLOB bound for long distances. We have numerically verified that, by optimizing system parameters, even for longer distances than shown on the graph, we can obtain positive key rates, albeit very low, for QS-assisted systems. The post-selection mechanism in the QS seems to be the key to obtaining positive key rates at long distances. At such distances, the channel loss naturally prepares low-intensity inputs to the QS, which allows us to use larger values of $V_A$, as shown in Table~\ref{table}. That would also enable us to use higher gains without necessarily increasing the QS noise. A higher-than unity gain for the post-selected states would then offer a better signal-to-noise ratio at long distances, which allows us to achieve positive secret key rates at longer distances than can otherwise be achieved for a no-QS system. 

Figure~\ref{fig:rates} also shows that our QS-amplified system cannot beat the existing upper bound for repeaterless systems \cite{Pirandola_PLOB17}. This agrees with the fact that any postprocessing at the receiver side does not change the repeaterless nature of the link, even though a form of amplification is in use. But, it will be interesting to see if, based on the above results, we can assess the practicality of the proposed CV repeater setups as in  \cite{Dias_Ralph_CVQRs}. On the positive side, we can see that there exists a regime of operation where the slope of the QS-based curves offer a square root advantage as needed in repeater systems. On the downside, however, this behaviour only appears in a limited range of distance, and only up to a maximum value of excess noise. In our simulations, we were not able to obtain any positive secret key rates at $\varepsilon_{\rm tm} = 0.06$, or higher. It seems that once the starting distance at which QS-based curves offer positive key rates lie above the maximum security distance for no-QS systems, it is no longer possible to get a positive key rate for QS-assisted systems. This may suggest that similar limitations might affect the suitability of CV repeater systems for QKD applications, which needs further investigation.

\begin{table}[t]
	\centering
	\caption{Optimized values for modulation variance and amplification gain at zero excess noise for the QS-based system.}      
	\begin{tabular}{|c|c|c|} \hline
		{Distance (km)} & {Optimized $V_A$} & {Optimized gain, $g$ } \\ \hline
		0 & 0.05 & 1.00  \\ \hline
		100 & 0.8 & 1.36  \\ \hline
		200 & 3.5 & 2.38 \\ \hline
		300 & 11.5 & 4.36  \\ \hline
		400 & 12.5 & 14.1  \\ \hline
		500 & 13.5 & 100  \\ \hline
	\end{tabular}
	\label{table} 
\end{table}

\section{Conclusion}
\label{sec:conclusion}
In this work, we studied the performance of the GG02 protocol where the received signal was amplified by a quantum scissor. We first obtained the exact output state and success probability of the QS under study, which was later used in calculating the secret key generation rate of the system. 
We showed that the QS would turn a Gaussian input state into a non-Gaussian one. That would make the conventional techniques to estimating the key rate not directly applicable to our case. 
We instead directly calculated the mutual information by working out the probability distribution function of the quadratures after the QS. Also, in order to calculate the leaked information to Eve, we obtained the exact covariance matrix of the bipartite state shared between sender and receiver labs in the particular case of a Gaussian attack. We then found the Holevo information corresponding to a Gaussian shared output state with the same covariance matrix, which gives an upper bound for the Holevo term in the case considered. We optimized the key rate over input modulation variance and amplification gain. Our results showed that, for a certain range of excess noise, the QS-enhanced system could reach longer distances than the no-QS system.  

There are certain practical aspects that one should consider before using quantum scissors in CV\:QKD. One assumption that we make throughout our paper is that on-demand single-photon sources are available for our scheme. There are two practical issues, in this regard, that affect the performance of the QS-based system. The first is the rate at which single-photons are generated. The success rate of such sources directly affect the key rate achievable. Secondly, we should be cautious about the purity of the single-photon source output. Multiple-photon components, in particular, could be damaging to the performance of the QS. The good news is that the current available technology for quantum-dot sources has made a substantial progress to meet both above requirements. In particular, quantum dot sources with efficiencies over 80\% and second-order coherence values $<0.004$ have already been demonstrated \cite{QDot_entg_low_g2:NatPhot2014,Senellart_qdots:NatPhot2017}. The second issue is the reliance on single-photon detectors, which will make CV\:QKD systems, in terms of requirements, similar to their discrete-variable counterparts. But, paying such prices may be unavoidable if one wants to have long-distance CV\:QKD and/or CV repeaters. Our study would, in particular, be highly relevant to analyzing the performance of recently proposed CV quantum repeaters \cite{Dias_Ralph_CVQRs}, which rely on a similar building block. Moreover, one should note that all these additional equipment are at the receiver end of the CV-QKD link, which is often located at a network node, shared among many users. This can bring the total cost per user down to a reasonable value when the system is in  widespread use.

We conclude by pointing out two additional remarks. First, note that, while the original NLA proposal by Ralph and Lund relies on multiple QS modules, in our scheme, we find using one QS is optimal as it minimizes the noise while we can adjust the signal level by optimizing the modulation variance. This also agrees with the results reported in  \cite{Guha_CV_Repeater}, where they have shown that the reverse coherent information \cite{Pirandola:RevKeyCap2009,Garcia-Patron:RevCohInf2009} is maximum when one QS is used. Secondly, one may wonder about the similarities versus differences of an alternative approach to improving the rate-versus-distance behavior in CV\:QKD based on fighting noise by adding trusted noise \cite{Pirandola:RevKeyCap2009,PRL_trustednoise1,NCOM_trustednoise2} with the NLA solution. While, in our QS-based system, there are some elements of controlled noise by injecting the vacuum state into the QS module we believe that the key advantage of using a QS is in its underlying {\em post-selected} output. It will remain as an open question for future research to determine which of the two solutions are more effective in different scenarios, and if their impact can be combined to come up with more loss-resilient CV\:QKD implementations.

\begin{acknowledgments}
The authors acknowledge partial support from the White Rose Research Studentship and the UK EPSRC Grant No. EP/M013472/1. S.P. would like to acknowledge funding from the European Union’s Horizon 2020 research and innovation program under grant agreement No. 820466 (Continuous Variable Quantum Communications, `CiViQ'). All data generated in this paper can be reproduced by the provided methodology and equations.
\end{acknowledgments}

%%%%%%%%%%%%%%%%%%%%%%%%%%%%%%%%%%%%%
\bibliography{biblio_ghalaii} 

\begin{thebibliography}{51}%
\makeatletter
\providecommand \@ifxundefined [1]{%
 \@ifx{#1\undefined}
}%
\providecommand \@ifnum [1]{%
 \ifnum #1\expandafter \@firstoftwo
 \else \expandafter \@secondoftwo
 \fi
}%
\providecommand \@ifx [1]{%
 \ifx #1\expandafter \@firstoftwo
 \else \expandafter \@secondoftwo
 \fi
}%
\providecommand \natexlab [1]{#1}%
\providecommand \enquote  [1]{``#1''}%
\providecommand \bibnamefont  [1]{#1}%
\providecommand \bibfnamefont [1]{#1}%
\providecommand \citenamefont [1]{#1}%
\providecommand \href@noop [0]{\@secondoftwo}%
\providecommand \href [0]{\begingroup \@sanitize@url \@href}%
\providecommand \@href[1]{\@@startlink{#1}\@@href}%
\providecommand \@@href[1]{\endgroup#1\@@endlink}%
\providecommand \@sanitize@url [0]{\catcode `\\12\catcode `\$12\catcode
  `\&12\catcode `\#12\catcode `\^12\catcode `\_12\catcode `\%12\relax}%
\providecommand \@@startlink[1]{}%
\providecommand \@@endlink[0]{}%
\providecommand \url  [0]{\begingroup\@sanitize@url \@url }%
\providecommand \@url [1]{\endgroup\@href {#1}{\urlprefix }}%
\providecommand \urlprefix  [0]{URL }%
\providecommand \Eprint [0]{\href }%
\providecommand \doibase [0]{http://dx.doi.org/}%
\providecommand \selectlanguage [0]{\@gobble}%
\providecommand \bibinfo  [0]{\@secondoftwo}%
\providecommand \bibfield  [0]{\@secondoftwo}%
\providecommand \translation [1]{[#1]}%
\providecommand \BibitemOpen [0]{}%
\providecommand \bibitemStop [0]{}%
\providecommand \bibitemNoStop [0]{.\EOS\space}%
\providecommand \EOS [0]{\spacefactor3000\relax}%
\providecommand \BibitemShut  [1]{\csname bibitem#1\endcsname}%
\let\auto@bib@innerbib\@empty
%</preamble>
\bibitem [{\citenamefont {Pirandola}\ \emph {et~al.}(2019)\citenamefont
  {Pirandola}, \citenamefont {Andersen}, \citenamefont {Banchi}, \citenamefont
  {Berta}, \citenamefont {Bunandar}, \citenamefont {Colbeck}, \citenamefont
  {Englund}, \citenamefont {Gehring}, \citenamefont {Lupo}, \citenamefont
  {Ottaviani}, \citenamefont {Pereira}, \citenamefont {Razavi}, \citenamefont
  {Shaari}, \citenamefont {Tomamichel}, \citenamefont {Usenko}, \citenamefont
  {Vallone}, \citenamefont {Villoresi},\ and\ \citenamefont
  {Wallden}}]{Pirandola:RevQKD2019}%
  \BibitemOpen
  \bibfield  {author} {\bibinfo {author} {\bibfnamefont {S.}~\bibnamefont
  {Pirandola}}, \bibinfo {author} {\bibfnamefont {U.~L.}\ \bibnamefont
  {Andersen}}, \bibinfo {author} {\bibfnamefont {L.}~\bibnamefont {Banchi}},
  \bibinfo {author} {\bibfnamefont {M.}~\bibnamefont {Berta}}, \bibinfo
  {author} {\bibfnamefont {D.}~\bibnamefont {Bunandar}}, \bibinfo {author}
  {\bibfnamefont {R.}~\bibnamefont {Colbeck}}, \bibinfo {author} {\bibfnamefont
  {D.}~\bibnamefont {Englund}}, \bibinfo {author} {\bibfnamefont
  {T.}~\bibnamefont {Gehring}}, \bibinfo {author} {\bibfnamefont
  {C.}~\bibnamefont {Lupo}}, \bibinfo {author} {\bibfnamefont {C.}~\bibnamefont
  {Ottaviani}}, \bibinfo {author} {\bibfnamefont {J.}~\bibnamefont {Pereira}},
  \bibinfo {author} {\bibfnamefont {M.}~\bibnamefont {Razavi}}, \bibinfo
  {author} {\bibfnamefont {J.~S.}\ \bibnamefont {Shaari}}, \bibinfo {author}
  {\bibfnamefont {M.}~\bibnamefont {Tomamichel}}, \bibinfo {author}
  {\bibfnamefont {V.~C.}\ \bibnamefont {Usenko}}, \bibinfo {author}
  {\bibfnamefont {G.}~\bibnamefont {Vallone}}, \bibinfo {author} {\bibfnamefont
  {P.}~\bibnamefont {Villoresi}}, \ and\ \bibinfo {author} {\bibfnamefont
  {P.}~\bibnamefont {Wallden}},\ }\href@noop {} {\bibfield  {journal} {\bibinfo
   {journal} {arXiv:1906.01645}\ } (\bibinfo {year} {2019})}\BibitemShut
  {NoStop}%
\bibitem [{\citenamefont {Bennett}\ and\ \citenamefont
  {Brassard}(1984)}]{Bennett_BB84}%
  \BibitemOpen
  \bibfield  {author} {\bibinfo {author} {\bibfnamefont {C.~H.}\ \bibnamefont
  {Bennett}}\ and\ \bibinfo {author} {\bibfnamefont {G.}~\bibnamefont
  {Brassard}},\ }in\ \href@noop {} {\emph {\bibinfo {booktitle} {Proceedings of
  IEEE International Conference on Computers, Systems, and Signal
  Processing}}}\ (\bibinfo {address} {Bangalore, India},\ \bibinfo {year}
  {1984})\ pp.\ \bibinfo {pages} {175--179}\BibitemShut {NoStop}%
\bibitem [{\citenamefont {Ekert}(1991)}]{Ekert_1991}%
  \BibitemOpen
  \bibfield  {author} {\bibinfo {author} {\bibfnamefont {A.~K.}\ \bibnamefont
  {Ekert}},\ }\href@noop {} {\bibfield  {journal} {\bibinfo  {journal} {Phys.
  Rev. Lett.}\ }\textbf {\bibinfo {volume} {67}},\ \bibinfo {pages} {661}
  (\bibinfo {year} {1991})}\BibitemShut {NoStop}%
\bibitem [{\citenamefont {Gisin}\ \emph {et~al.}(2002)\citenamefont {Gisin},
  \citenamefont {Ribordy}, \citenamefont {Tittel},\ and\ \citenamefont
  {Zbinden}}]{Gisin_QCrypRev_2002}%
  \BibitemOpen
  \bibfield  {author} {\bibinfo {author} {\bibfnamefont {N.}~\bibnamefont
  {Gisin}}, \bibinfo {author} {\bibfnamefont {G.}~\bibnamefont {Ribordy}},
  \bibinfo {author} {\bibfnamefont {W.}~\bibnamefont {Tittel}}, \ and\ \bibinfo
  {author} {\bibfnamefont {H.}~\bibnamefont {Zbinden}},\ }\href@noop {}
  {\bibfield  {journal} {\bibinfo  {journal} {Rev. Mod. Phys.}\ }\textbf
  {\bibinfo {volume} {74}},\ \bibinfo {pages} {145} (\bibinfo {year}
  {2002})}\BibitemShut {NoStop}%
\bibitem [{\citenamefont {Scarani}\ \emph {et~al.}(2009)\citenamefont
  {Scarani}, \citenamefont {Bechmann-Pasquinucci}, \citenamefont {Cerf},
  \citenamefont {Du\ifmmode~\check{s}\else \v{s}\fi{}ek}, \citenamefont
  {L\"utkenhaus},\ and\ \citenamefont {Peev}}]{Scarani_QKDRev_2009}%
  \BibitemOpen
  \bibfield  {author} {\bibinfo {author} {\bibfnamefont {V.}~\bibnamefont
  {Scarani}}, \bibinfo {author} {\bibfnamefont {H.}~\bibnamefont
  {Bechmann-Pasquinucci}}, \bibinfo {author} {\bibfnamefont {N.~J.}\
  \bibnamefont {Cerf}}, \bibinfo {author} {\bibfnamefont {M.}~\bibnamefont
  {Du\ifmmode~\check{s}\else \v{s}\fi{}ek}}, \bibinfo {author} {\bibfnamefont
  {N.}~\bibnamefont {L\"utkenhaus}}, \ and\ \bibinfo {author} {\bibfnamefont
  {M.}~\bibnamefont {Peev}},\ }\href@noop {} {\bibfield  {journal} {\bibinfo
  {journal} {Rev. Mod. Phys.}\ }\textbf {\bibinfo {volume} {81}},\ \bibinfo
  {pages} {1301} (\bibinfo {year} {2009})}\BibitemShut {NoStop}%
\bibitem [{\citenamefont {Grosshans}\ and\ \citenamefont
  {Grangier}(2002)}]{Grosshans_GG02_PRL}%
  \BibitemOpen
  \bibfield  {author} {\bibinfo {author} {\bibfnamefont {F.}~\bibnamefont
  {Grosshans}}\ and\ \bibinfo {author} {\bibfnamefont {P.}~\bibnamefont
  {Grangier}},\ }\href@noop {} {\bibfield  {journal} {\bibinfo  {journal}
  {Phys. Rev. Lett.}\ }\textbf {\bibinfo {volume} {88}},\ \bibinfo {pages}
  {057902} (\bibinfo {year} {2002})}\BibitemShut {NoStop}%
\bibitem [{\citenamefont {Grosshans}\ \emph {et~al.}(2003)\citenamefont
  {Grosshans}, \citenamefont {Van~Assche}, \citenamefont {Wenger},
  \citenamefont {Brouri}, \citenamefont {Cerf},\ and\ \citenamefont
  {Grangier}}]{Grosshans_GG02_Nature}%
  \BibitemOpen
  \bibfield  {author} {\bibinfo {author} {\bibfnamefont {F.}~\bibnamefont
  {Grosshans}}, \bibinfo {author} {\bibfnamefont {G.}~\bibnamefont
  {Van~Assche}}, \bibinfo {author} {\bibfnamefont {J.}~\bibnamefont {Wenger}},
  \bibinfo {author} {\bibfnamefont {R.}~\bibnamefont {Brouri}}, \bibinfo
  {author} {\bibfnamefont {N.~J.}\ \bibnamefont {Cerf}}, \ and\ \bibinfo
  {author} {\bibfnamefont {P.}~\bibnamefont {Grangier}},\ }\href {\doibase
  10.1038/nature01289} {\bibfield  {journal} {\bibinfo  {journal} {Nature}\
  }\textbf {\bibinfo {volume} {421}},\ \bibinfo {pages} {238} (\bibinfo {year}
  {2003})}\BibitemShut {NoStop}%
\bibitem [{\citenamefont {Braunstein}\ and\ \citenamefont {van
  Loock}(2005)}]{Braunstein_QICVRev_2005}%
  \BibitemOpen
  \bibfield  {author} {\bibinfo {author} {\bibfnamefont {S.~L.}\ \bibnamefont
  {Braunstein}}\ and\ \bibinfo {author} {\bibfnamefont {P.}~\bibnamefont {van
  Loock}},\ }\href@noop {} {\bibfield  {journal} {\bibinfo  {journal} {Rev.
  Mod. Phys.}\ }\textbf {\bibinfo {volume} {77}},\ \bibinfo {pages} {513}
  (\bibinfo {year} {2005})}\BibitemShut {NoStop}%
\bibitem [{\citenamefont {Weedbrook}\ \emph {et~al.}(2012)\citenamefont
  {Weedbrook}, \citenamefont {Pirandola}, \citenamefont {Garc\'{\i}a-Patr\'on},
  \citenamefont {Cerf}, \citenamefont {Ralph}, \citenamefont {Shapiro},\ and\
  \citenamefont {Lloyd}}]{Weedbrook_GaussQIRev_2012}%
  \BibitemOpen
  \bibfield  {author} {\bibinfo {author} {\bibfnamefont {C.}~\bibnamefont
  {Weedbrook}}, \bibinfo {author} {\bibfnamefont {S.}~\bibnamefont
  {Pirandola}}, \bibinfo {author} {\bibfnamefont {R.}~\bibnamefont
  {Garc\'{\i}a-Patr\'on}}, \bibinfo {author} {\bibfnamefont {N.~J.}\
  \bibnamefont {Cerf}}, \bibinfo {author} {\bibfnamefont {T.~C.}\ \bibnamefont
  {Ralph}}, \bibinfo {author} {\bibfnamefont {J.~H.}\ \bibnamefont {Shapiro}},
  \ and\ \bibinfo {author} {\bibfnamefont {S.}~\bibnamefont {Lloyd}},\
  }\href@noop {} {\bibfield  {journal} {\bibinfo  {journal} {Rev. Mod. Phys.}\
  }\textbf {\bibinfo {volume} {84}},\ \bibinfo {pages} {621} (\bibinfo {year}
  {2012})}\BibitemShut {NoStop}%
\bibitem [{\citenamefont {Jouguet}\ \emph {et~al.}()\citenamefont {Jouguet},
  \citenamefont {Kunz-Jacques}, \citenamefont {Leverrier}, \citenamefont
  {Grangier},\ and\ \citenamefont {Diamanti}}]{Jouguet_LD_CVQKD_NatPhoton}%
  \BibitemOpen
  \bibfield  {author} {\bibinfo {author} {\bibfnamefont {P.}~\bibnamefont
  {Jouguet}}, \bibinfo {author} {\bibfnamefont {S.}~\bibnamefont
  {Kunz-Jacques}}, \bibinfo {author} {\bibfnamefont {A.}~\bibnamefont
  {Leverrier}}, \bibinfo {author} {\bibfnamefont {P.}~\bibnamefont {Grangier}},
  \ and\ \bibinfo {author} {\bibfnamefont {E.}~\bibnamefont {Diamanti}},\
  }\href@noop {} {\bibfield  {journal} {\bibinfo  {journal} {Nat. Photon.}\
  }\textbf {\bibinfo {volume} {7}},\ \bibinfo {pages} {378}}\BibitemShut
  {NoStop}%
\bibitem [{\citenamefont {Pirandola}\ \emph {et~al.}(2015)\citenamefont
  {Pirandola}, \citenamefont {Ottaviani}, \citenamefont {Spedalieri},
  \citenamefont {Weedbrook}, \citenamefont {Braunstein}, \citenamefont {Lloyd},
  \citenamefont {Gehring}, \citenamefont {Jacobsen},\ and\ \citenamefont
  {Andersen}}]{Pirandola2015}%
  \BibitemOpen
  \bibfield  {author} {\bibinfo {author} {\bibfnamefont {S.}~\bibnamefont
  {Pirandola}}, \bibinfo {author} {\bibfnamefont {C.}~\bibnamefont
  {Ottaviani}}, \bibinfo {author} {\bibfnamefont {G.}~\bibnamefont
  {Spedalieri}}, \bibinfo {author} {\bibfnamefont {C.}~\bibnamefont
  {Weedbrook}}, \bibinfo {author} {\bibfnamefont {S.~L.}\ \bibnamefont
  {Braunstein}}, \bibinfo {author} {\bibfnamefont {S.}~\bibnamefont {Lloyd}},
  \bibinfo {author} {\bibfnamefont {T.}~\bibnamefont {Gehring}}, \bibinfo
  {author} {\bibfnamefont {C.~S.}\ \bibnamefont {Jacobsen}}, \ and\ \bibinfo
  {author} {\bibfnamefont {U.~L.}\ \bibnamefont {Andersen}},\ }\href {\doibase
  10.1038/nphoton.2015.83} {\bibfield  {journal} {\bibinfo  {journal} {Nat.
  Photon.}\ }\textbf {\bibinfo {volume} {9}},\ \bibinfo {pages} {397} (\bibinfo
  {year} {2015})}\BibitemShut {NoStop}%
\bibitem [{\citenamefont {Hirano}\ \emph {et~al.}(2003)\citenamefont {Hirano},
  \citenamefont {Yamanaka}, \citenamefont {Ashikaga}, \citenamefont {Konishi},\
  and\ \citenamefont {Namiki}}]{Hirano_PulsedHOM}%
  \BibitemOpen
  \bibfield  {author} {\bibinfo {author} {\bibfnamefont {T.}~\bibnamefont
  {Hirano}}, \bibinfo {author} {\bibfnamefont {H.}~\bibnamefont {Yamanaka}},
  \bibinfo {author} {\bibfnamefont {M.}~\bibnamefont {Ashikaga}}, \bibinfo
  {author} {\bibfnamefont {T.}~\bibnamefont {Konishi}}, \ and\ \bibinfo
  {author} {\bibfnamefont {R.}~\bibnamefont {Namiki}},\ }\href@noop {}
  {\bibfield  {journal} {\bibinfo  {journal} {Phys. Rev. A}\ }\textbf {\bibinfo
  {volume} {68}},\ \bibinfo {pages} {042331} (\bibinfo {year}
  {2003})}\BibitemShut {NoStop}%
\bibitem [{\citenamefont {Yonezawa}\ \emph {et~al.}(2007)\citenamefont
  {Yonezawa}, \citenamefont {Braunstein},\ and\ \citenamefont
  {Furusawa}}]{Yonezawa_QTele_Exp}%
  \BibitemOpen
  \bibfield  {author} {\bibinfo {author} {\bibfnamefont {H.}~\bibnamefont
  {Yonezawa}}, \bibinfo {author} {\bibfnamefont {S.~L.}\ \bibnamefont
  {Braunstein}}, \ and\ \bibinfo {author} {\bibfnamefont {A.}~\bibnamefont
  {Furusawa}},\ }\href@noop {} {\bibfield  {journal} {\bibinfo  {journal}
  {Phys. Rev. Lett.}\ }\textbf {\bibinfo {volume} {99}},\ \bibinfo {pages}
  {110503} (\bibinfo {year} {2007})}\BibitemShut {NoStop}%
\bibitem [{\citenamefont {Yokoyama}\ \emph {et~al.}(2013)\citenamefont
  {Yokoyama}, \citenamefont {Ukai}, \citenamefont {Armstrong}, \citenamefont
  {Sornphiphatphong}, \citenamefont {Kaji}, \citenamefont {Suzuki},
  \citenamefont {ichi Yoshikawa}, \citenamefont {Yonezawa}, \citenamefont
  {Menicucci},\ and\ \citenamefont {Furusawa}}]{Yokoyama_NatPhoton}%
  \BibitemOpen
  \bibfield  {author} {\bibinfo {author} {\bibfnamefont {S.}~\bibnamefont
  {Yokoyama}}, \bibinfo {author} {\bibfnamefont {R.}~\bibnamefont {Ukai}},
  \bibinfo {author} {\bibfnamefont {S.~C.}\ \bibnamefont {Armstrong}}, \bibinfo
  {author} {\bibfnamefont {C.}~\bibnamefont {Sornphiphatphong}}, \bibinfo
  {author} {\bibfnamefont {T.}~\bibnamefont {Kaji}}, \bibinfo {author}
  {\bibfnamefont {S.}~\bibnamefont {Suzuki}}, \bibinfo {author} {\bibfnamefont
  {J.}~\bibnamefont {ichi Yoshikawa}}, \bibinfo {author} {\bibfnamefont
  {H.}~\bibnamefont {Yonezawa}}, \bibinfo {author} {\bibfnamefont {N.~C.}\
  \bibnamefont {Menicucci}}, \ and\ \bibinfo {author} {\bibfnamefont
  {A.}~\bibnamefont {Furusawa}},\ }\href@noop {} {\bibfield  {journal}
  {\bibinfo  {journal} {Nat. Photon.}\ }\textbf {\bibinfo {volume} {7}},\
  \bibinfo {pages} {982} (\bibinfo {year} {2013})}\BibitemShut {NoStop}%
\bibitem [{\citenamefont {Jouguet}\ \emph {et~al.}(2011)\citenamefont
  {Jouguet}, \citenamefont {Kunz-Jacques},\ and\ \citenamefont
  {Leverrier}}]{Jouguet_LD_CVQKD_PRA}%
  \BibitemOpen
  \bibfield  {author} {\bibinfo {author} {\bibfnamefont {P.}~\bibnamefont
  {Jouguet}}, \bibinfo {author} {\bibfnamefont {S.}~\bibnamefont
  {Kunz-Jacques}}, \ and\ \bibinfo {author} {\bibfnamefont {A.}~\bibnamefont
  {Leverrier}},\ }\href@noop {} {\bibfield  {journal} {\bibinfo  {journal}
  {Phys. Rev. A}\ }\textbf {\bibinfo {volume} {84}},\ \bibinfo {pages} {062317}
  (\bibinfo {year} {2011})}\BibitemShut {NoStop}%
\bibitem [{\citenamefont {Blandino}\ \emph {et~al.}(2012)\citenamefont
  {Blandino}, \citenamefont {Leverrier}, \citenamefont {Barbieri},
  \citenamefont {Etesse}, \citenamefont {Grangier},\ and\ \citenamefont
  {Tualle-Brouri}}]{Blandino_CVQKD_idealNLA}%
  \BibitemOpen
  \bibfield  {author} {\bibinfo {author} {\bibfnamefont {R.}~\bibnamefont
  {Blandino}}, \bibinfo {author} {\bibfnamefont {A.}~\bibnamefont {Leverrier}},
  \bibinfo {author} {\bibfnamefont {M.}~\bibnamefont {Barbieri}}, \bibinfo
  {author} {\bibfnamefont {J.}~\bibnamefont {Etesse}}, \bibinfo {author}
  {\bibfnamefont {P.}~\bibnamefont {Grangier}}, \ and\ \bibinfo {author}
  {\bibfnamefont {R.}~\bibnamefont {Tualle-Brouri}},\ }\href@noop {} {\bibfield
   {journal} {\bibinfo  {journal} {Phys. Rev. A}\ }\textbf {\bibinfo {volume}
  {86}},\ \bibinfo {pages} {012327} (\bibinfo {year} {2012})}\BibitemShut
  {NoStop}%
\bibitem [{\citenamefont {Zhang}\ \emph {et~al.}()\citenamefont {Zhang},
  \citenamefont {Li}, \citenamefont {Weedbrook}, \citenamefont {Yu},
  \citenamefont {Gu}, \citenamefont {Sun}, \citenamefont {Peng},\ and\
  \citenamefont {Guo}}]{Zhang_2Way_CVQKD_Amp}%
  \BibitemOpen
  \bibfield  {author} {\bibinfo {author} {\bibfnamefont {Y.-C.}\ \bibnamefont
  {Zhang}}, \bibinfo {author} {\bibfnamefont {Z.}~\bibnamefont {Li}}, \bibinfo
  {author} {\bibfnamefont {C.}~\bibnamefont {Weedbrook}}, \bibinfo {author}
  {\bibfnamefont {S.}~\bibnamefont {Yu}}, \bibinfo {author} {\bibfnamefont
  {W.}~\bibnamefont {Gu}}, \bibinfo {author} {\bibfnamefont {M.}~\bibnamefont
  {Sun}}, \bibinfo {author} {\bibfnamefont {X.}~\bibnamefont {Peng}}, \ and\
  \bibinfo {author} {\bibfnamefont {H.}~\bibnamefont {Guo}},\ }\href@noop {}
  {\bibfield  {journal} {\bibinfo  {journal} {J. Phys. B: At. Mol. Opt. Phys.}\
  }\textbf {\bibinfo {volume} {47}},\ \bibinfo {pages} {035501}}\BibitemShut
  {NoStop}%
\bibitem [{\citenamefont {Pandey}\ \emph {et~al.}(2013)\citenamefont {Pandey},
  \citenamefont {Jiang}, \citenamefont {Combes},\ and\ \citenamefont
  {Caves}}]{Pandey_Qlimits_Amp}%
  \BibitemOpen
  \bibfield  {author} {\bibinfo {author} {\bibfnamefont {S.}~\bibnamefont
  {Pandey}}, \bibinfo {author} {\bibfnamefont {Z.}~\bibnamefont {Jiang}},
  \bibinfo {author} {\bibfnamefont {J.}~\bibnamefont {Combes}}, \ and\ \bibinfo
  {author} {\bibfnamefont {C.~M.}\ \bibnamefont {Caves}},\ }\href@noop {}
  {\bibfield  {journal} {\bibinfo  {journal} {Phys. Rev. A}\ }\textbf {\bibinfo
  {volume} {88}},\ \bibinfo {pages} {033852} (\bibinfo {year}
  {2013})}\BibitemShut {NoStop}%
\bibitem [{\citenamefont {Pegg}\ \emph {et~al.}(1998)\citenamefont {Pegg},
  \citenamefont {Phillips},\ and\ \citenamefont {Barnett}}]{Pegg_QSs}%
  \BibitemOpen
  \bibfield  {author} {\bibinfo {author} {\bibfnamefont {D.~T.}\ \bibnamefont
  {Pegg}}, \bibinfo {author} {\bibfnamefont {L.~S.}\ \bibnamefont {Phillips}},
  \ and\ \bibinfo {author} {\bibfnamefont {S.~M.}\ \bibnamefont {Barnett}},\
  }\href@noop {} {\bibfield  {journal} {\bibinfo  {journal} {Phys. Rev. Lett.}\
  }\textbf {\bibinfo {volume} {81}},\ \bibinfo {pages} {1604} (\bibinfo {year}
  {1998})}\BibitemShut {NoStop}%
\bibitem [{\citenamefont {Ralph}\ and\ \citenamefont
  {Lund}(2009)}]{Ralph_Lund_QSNLA}%
  \BibitemOpen
  \bibfield  {author} {\bibinfo {author} {\bibfnamefont {T.~C.}\ \bibnamefont
  {Ralph}}\ and\ \bibinfo {author} {\bibfnamefont {A.~P.}\ \bibnamefont
  {Lund}},\ }\href@noop {} {\bibfield  {journal} {\bibinfo  {journal} {AIP
  Conference Proceedings}\ }\textbf {\bibinfo {volume} {1110}},\ \bibinfo
  {pages} {155} (\bibinfo {year} {2009})}\BibitemShut {NoStop}%
\bibitem [{\citenamefont {Eleftheriadou}\ \emph {et~al.}(2013)\citenamefont
  {Eleftheriadou}, \citenamefont {Barnett},\ and\ \citenamefont
  {Jeffers}}]{Eleftheriadou_QAmp}%
  \BibitemOpen
  \bibfield  {author} {\bibinfo {author} {\bibfnamefont {E.}~\bibnamefont
  {Eleftheriadou}}, \bibinfo {author} {\bibfnamefont {S.~M.}\ \bibnamefont
  {Barnett}}, \ and\ \bibinfo {author} {\bibfnamefont {J.}~\bibnamefont
  {Jeffers}},\ }\href@noop {} {\bibfield  {journal} {\bibinfo  {journal} {Phys.
  Rev. Lett.}\ }\textbf {\bibinfo {volume} {111}},\ \bibinfo {pages} {213601}
  (\bibinfo {year} {2013})}\BibitemShut {NoStop}%
\bibitem [{\citenamefont {Fiur\'a\ifmmode~\check{s}\else
  \v{s}\fi{}ek}(2009)}]{Fiurasek}%
  \BibitemOpen
  \bibfield  {author} {\bibinfo {author} {\bibfnamefont {J.}~\bibnamefont
  {Fiur\'a\ifmmode~\check{s}\else \v{s}\fi{}ek}},\ }\href@noop {} {\bibfield
  {journal} {\bibinfo  {journal} {Phys. Rev. A}\ }\textbf {\bibinfo {volume}
  {80}},\ \bibinfo {pages} {053822} (\bibinfo {year} {2009})}\BibitemShut
  {NoStop}%
\bibitem [{\citenamefont {Xiang}\ \emph {et~al.}(2010)\citenamefont {Xiang},
  \citenamefont {Ralph}, \citenamefont {Lund}, \citenamefont {Walk},\ and\
  \citenamefont {Pryde}}]{Xiang_NatPhys}%
  \BibitemOpen
  \bibfield  {author} {\bibinfo {author} {\bibfnamefont {G.~Y.}\ \bibnamefont
  {Xiang}}, \bibinfo {author} {\bibfnamefont {T.~C.}\ \bibnamefont {Ralph}},
  \bibinfo {author} {\bibfnamefont {A.~P.}\ \bibnamefont {Lund}}, \bibinfo
  {author} {\bibfnamefont {N.}~\bibnamefont {Walk}}, \ and\ \bibinfo {author}
  {\bibfnamefont {G.~J.}\ \bibnamefont {Pryde}},\ }\href {\doibase
  10.1038/nphoton.2010.35} {\bibfield  {journal} {\bibinfo  {journal} {Nat.
  Photon.}\ }\textbf {\bibinfo {volume} {4}},\ \bibinfo {pages} {316} (\bibinfo
  {year} {2010})}\BibitemShut {NoStop}%
\bibitem [{\citenamefont {Ferreyrol}\ \emph {et~al.}(2010)\citenamefont
  {Ferreyrol}, \citenamefont {Barbieri}, \citenamefont {Blandino},
  \citenamefont {Fossier}, \citenamefont {Tualle-Brouri},\ and\ \citenamefont
  {Grangier}}]{Ferreyrol_ImpNLA_PRL}%
  \BibitemOpen
  \bibfield  {author} {\bibinfo {author} {\bibfnamefont {F.}~\bibnamefont
  {Ferreyrol}}, \bibinfo {author} {\bibfnamefont {M.}~\bibnamefont {Barbieri}},
  \bibinfo {author} {\bibfnamefont {R.}~\bibnamefont {Blandino}}, \bibinfo
  {author} {\bibfnamefont {S.}~\bibnamefont {Fossier}}, \bibinfo {author}
  {\bibfnamefont {R.}~\bibnamefont {Tualle-Brouri}}, \ and\ \bibinfo {author}
  {\bibfnamefont {P.}~\bibnamefont {Grangier}},\ }\href@noop {} {\bibfield
  {journal} {\bibinfo  {journal} {Phys. Rev. Lett.}\ }\textbf {\bibinfo
  {volume} {104}},\ \bibinfo {pages} {123603} (\bibinfo {year}
  {2010})}\BibitemShut {NoStop}%
\bibitem [{\citenamefont {Donaldson}\ \emph {et~al.}(2015)\citenamefont
  {Donaldson}, \citenamefont {Collins}, \citenamefont {Eleftheriadou},
  \citenamefont {Barnett}, \citenamefont {Jeffers},\ and\ \citenamefont
  {Buller}}]{Donaldson_ImpNLA_PRL}%
  \BibitemOpen
  \bibfield  {author} {\bibinfo {author} {\bibfnamefont {R.~J.}\ \bibnamefont
  {Donaldson}}, \bibinfo {author} {\bibfnamefont {R.~J.}\ \bibnamefont
  {Collins}}, \bibinfo {author} {\bibfnamefont {E.}~\bibnamefont
  {Eleftheriadou}}, \bibinfo {author} {\bibfnamefont {S.~M.}\ \bibnamefont
  {Barnett}}, \bibinfo {author} {\bibfnamefont {J.}~\bibnamefont {Jeffers}}, \
  and\ \bibinfo {author} {\bibfnamefont {G.~S.}\ \bibnamefont {Buller}},\
  }\href@noop {} {\bibfield  {journal} {\bibinfo  {journal} {Phys. Rev. Lett.}\
  }\textbf {\bibinfo {volume} {114}},\ \bibinfo {pages} {120505} (\bibinfo
  {year} {2015})}\BibitemShut {NoStop}%
\bibitem [{\citenamefont {Barbieri}\ \emph {et~al.}()\citenamefont {Barbieri},
  \citenamefont {Ferreyrol}, \citenamefont {Blandino}, \citenamefont
  {Tualle-Brouri},\ and\ \citenamefont {Grangier}}]{Barbieri_NLA_Exp_Rev}%
  \BibitemOpen
  \bibfield  {author} {\bibinfo {author} {\bibfnamefont {M.}~\bibnamefont
  {Barbieri}}, \bibinfo {author} {\bibfnamefont {F.}~\bibnamefont {Ferreyrol}},
  \bibinfo {author} {\bibfnamefont {R.}~\bibnamefont {Blandino}}, \bibinfo
  {author} {\bibfnamefont {R.}~\bibnamefont {Tualle-Brouri}}, \ and\ \bibinfo
  {author} {\bibfnamefont {P.}~\bibnamefont {Grangier}},\ }\href@noop {}
  {\bibfield  {journal} {\bibinfo  {journal} {Laser Phys. Lett.}\ }\textbf
  {\bibinfo {volume} {8}},\ \bibinfo {pages} {411}}\BibitemShut {NoStop}%
\bibitem [{\citenamefont {Dias}\ and\ \citenamefont
  {Ralph}(2017)}]{Dias_Ralph_CVQRs}%
  \BibitemOpen
  \bibfield  {author} {\bibinfo {author} {\bibfnamefont {J.}~\bibnamefont
  {Dias}}\ and\ \bibinfo {author} {\bibfnamefont {T.~C.}\ \bibnamefont
  {Ralph}},\ }\href@noop {} {\bibfield  {journal} {\bibinfo  {journal} {Phys.
  Rev. A}\ }\textbf {\bibinfo {volume} {95}},\ \bibinfo {pages} {022312}
  (\bibinfo {year} {2017})}\BibitemShut {NoStop}%
\bibitem [{\citenamefont {Furrer}\ and\ \citenamefont
  {Munro}(2018)}]{Furrer_Munro_CVQRs}%
  \BibitemOpen
  \bibfield  {author} {\bibinfo {author} {\bibfnamefont {F.}~\bibnamefont
  {Furrer}}\ and\ \bibinfo {author} {\bibfnamefont {W.~J.}\ \bibnamefont
  {Munro}},\ }\href@noop {} {\bibfield  {journal} {\bibinfo  {journal} {Phys.
  Rev. A}\ }\textbf {\bibinfo {volume} {98}},\ \bibinfo {pages} {032335}
  (\bibinfo {year} {2018})}\BibitemShut {NoStop}%
\bibitem [{\citenamefont {Seshadreesan}\ \emph {et~al.}()\citenamefont
  {Seshadreesan}, \citenamefont {Krovi},\ and\ \citenamefont
  {Guha}}]{Guha_CV_Repeater}%
  \BibitemOpen
  \bibfield  {author} {\bibinfo {author} {\bibfnamefont {K.~P.}\ \bibnamefont
  {Seshadreesan}}, \bibinfo {author} {\bibfnamefont {H.}~\bibnamefont {Krovi}},
  \ and\ \bibinfo {author} {\bibfnamefont {S.}~\bibnamefont {Guha}},\
  }\href@noop {} {\bibinfo  {journal} {arXiv:1811.12393}\ }\BibitemShut
  {NoStop}%
\bibitem [{\citenamefont {Fiur\'a\ifmmode~\check{s}\else \v{s}\fi{}ek}\ and\
  \citenamefont {Cerf}(2012)}]{Fiurasek_VirNLA}%
  \BibitemOpen
\bibfield  {journal} {  }\bibfield  {author} {\bibinfo {author} {\bibfnamefont
  {J.}~\bibnamefont {Fiur\'a\ifmmode~\check{s}\else \v{s}\fi{}ek}}\ and\
  \bibinfo {author} {\bibfnamefont {N.~J.}\ \bibnamefont {Cerf}},\ }\href@noop
  {} {\bibfield  {journal} {\bibinfo  {journal} {Phys. Rev. A}\ }\textbf
  {\bibinfo {volume} {86}},\ \bibinfo {pages} {060302} (\bibinfo {year}
  {2012})}\BibitemShut {NoStop}%
\bibitem [{\citenamefont {Walk}\ \emph {et~al.}(2013)\citenamefont {Walk},
  \citenamefont {Ralph}, \citenamefont {Symul},\ and\ \citenamefont
  {Lam}}]{Walk_CVQKD_postsel}%
  \BibitemOpen
  \bibfield  {author} {\bibinfo {author} {\bibfnamefont {N.}~\bibnamefont
  {Walk}}, \bibinfo {author} {\bibfnamefont {T.~C.}\ \bibnamefont {Ralph}},
  \bibinfo {author} {\bibfnamefont {T.}~\bibnamefont {Symul}}, \ and\ \bibinfo
  {author} {\bibfnamefont {P.~K.}\ \bibnamefont {Lam}},\ }\href@noop {}
  {\bibfield  {journal} {\bibinfo  {journal} {Phys. Rev. A}\ }\textbf {\bibinfo
  {volume} {87}},\ \bibinfo {pages} {020303} (\bibinfo {year}
  {2013})}\BibitemShut {NoStop}%
\bibitem [{\citenamefont {Chrzanowski}\ \emph {et~al.}(2014)\citenamefont
  {Chrzanowski}, \citenamefont {Walk}, \citenamefont {Assad}, \citenamefont
  {Janousek}, \citenamefont {Hosseini}, \citenamefont {Ralph}, \citenamefont
  {Symul},\ and\ \citenamefont {Lam}}]{Chrzanowski_MBNLA}%
  \BibitemOpen
  \bibfield  {author} {\bibinfo {author} {\bibfnamefont {H.~M.}\ \bibnamefont
  {Chrzanowski}}, \bibinfo {author} {\bibfnamefont {N.}~\bibnamefont {Walk}},
  \bibinfo {author} {\bibfnamefont {S.~M.}\ \bibnamefont {Assad}}, \bibinfo
  {author} {\bibfnamefont {J.}~\bibnamefont {Janousek}}, \bibinfo {author}
  {\bibfnamefont {S.}~\bibnamefont {Hosseini}}, \bibinfo {author}
  {\bibfnamefont {T.~C.}\ \bibnamefont {Ralph}}, \bibinfo {author}
  {\bibfnamefont {T.}~\bibnamefont {Symul}}, \ and\ \bibinfo {author}
  {\bibfnamefont {P.~K.}\ \bibnamefont {Lam}},\ }\href@noop {} {\bibfield
  {journal} {\bibinfo  {journal} {Nat. Photon.}\ }\textbf {\bibinfo {volume}
  {8}},\ \bibinfo {pages} {333} (\bibinfo {year} {2014})}\BibitemShut {NoStop}%
\bibitem [{\citenamefont {Zhao}\ \emph {et~al.}(2017)\citenamefont {Zhao},
  \citenamefont {Haw}, \citenamefont {Symul}, \citenamefont {Lam},\ and\
  \citenamefont {Assad}}]{Zhao_VirNLA}%
  \BibitemOpen
  \bibfield  {author} {\bibinfo {author} {\bibfnamefont {J.}~\bibnamefont
  {Zhao}}, \bibinfo {author} {\bibfnamefont {J.~Y.}\ \bibnamefont {Haw}},
  \bibinfo {author} {\bibfnamefont {T.}~\bibnamefont {Symul}}, \bibinfo
  {author} {\bibfnamefont {P.~K.}\ \bibnamefont {Lam}}, \ and\ \bibinfo
  {author} {\bibfnamefont {S.~M.}\ \bibnamefont {Assad}},\ }\href@noop {}
  {\bibfield  {journal} {\bibinfo  {journal} {Phys. Rev. A}\ }\textbf {\bibinfo
  {volume} {96}},\ \bibinfo {pages} {012319} (\bibinfo {year}
  {2017})}\BibitemShut {NoStop}%
\bibitem [{\citenamefont {Bernu}\ \emph {et~al.}()\citenamefont {Bernu},
  \citenamefont {Armstrong}, \citenamefont {Symul}, \citenamefont {Ralph},\
  and\ \citenamefont {Lam}}]{Bernu_NLA_EPR_dist}%
  \BibitemOpen
  \bibfield  {author} {\bibinfo {author} {\bibfnamefont {J.}~\bibnamefont
  {Bernu}}, \bibinfo {author} {\bibfnamefont {S.}~\bibnamefont {Armstrong}},
  \bibinfo {author} {\bibfnamefont {T.}~\bibnamefont {Symul}}, \bibinfo
  {author} {\bibfnamefont {T.~C.}\ \bibnamefont {Ralph}}, \ and\ \bibinfo
  {author} {\bibfnamefont {P.~K.}\ \bibnamefont {Lam}},\ }\href@noop {}
  {\bibfield  {journal} {\bibinfo  {journal} {Journal of Physics B: Atomic,
  Molecular and Optical Physics}\ }\textbf {\bibinfo {volume} {47}},\ \bibinfo
  {pages} {215503}}\BibitemShut {NoStop}%
\bibitem [{\citenamefont {Lodewyck}\ \emph {et~al.}(2007)\citenamefont
  {Lodewyck}, \citenamefont {Bloch}, \citenamefont {Garc\'{\i}a-Patr\'on},
  \citenamefont {Fossier}, \citenamefont {Karpov}, \citenamefont {Diamanti},
  \citenamefont {Debuisschert}, \citenamefont {Cerf}, \citenamefont
  {Tualle-Brouri}, \citenamefont {McLaughlin},\ and\ \citenamefont
  {Grangier}}]{Lodewyck_25kmCVQKD}%
  \BibitemOpen
  \bibfield  {author} {\bibinfo {author} {\bibfnamefont {J.}~\bibnamefont
  {Lodewyck}}, \bibinfo {author} {\bibfnamefont {M.}~\bibnamefont {Bloch}},
  \bibinfo {author} {\bibfnamefont {R.}~\bibnamefont {Garc\'{\i}a-Patr\'on}},
  \bibinfo {author} {\bibfnamefont {S.}~\bibnamefont {Fossier}}, \bibinfo
  {author} {\bibfnamefont {E.}~\bibnamefont {Karpov}}, \bibinfo {author}
  {\bibfnamefont {E.}~\bibnamefont {Diamanti}}, \bibinfo {author}
  {\bibfnamefont {T.}~\bibnamefont {Debuisschert}}, \bibinfo {author}
  {\bibfnamefont {N.~J.}\ \bibnamefont {Cerf}}, \bibinfo {author}
  {\bibfnamefont {R.}~\bibnamefont {Tualle-Brouri}}, \bibinfo {author}
  {\bibfnamefont {S.~W.}\ \bibnamefont {McLaughlin}}, \ and\ \bibinfo {author}
  {\bibfnamefont {P.}~\bibnamefont {Grangier}},\ }\href@noop {} {\bibfield
  {journal} {\bibinfo  {journal} {Phys. Rev. A}\ }\textbf {\bibinfo {volume}
  {76}},\ \bibinfo {pages} {042305} (\bibinfo {year} {2007})}\BibitemShut
  {NoStop}%
\bibitem [{\citenamefont {Kumar}\ \emph {et~al.}(2015)\citenamefont {Kumar},
  \citenamefont {Qin},\ and\ \citenamefont {Alléaume}}]{Kumar_CVQKDwithDWDM}%
  \BibitemOpen
  \bibfield  {author} {\bibinfo {author} {\bibfnamefont {R.}~\bibnamefont
  {Kumar}}, \bibinfo {author} {\bibfnamefont {H.}~\bibnamefont {Qin}}, \ and\
  \bibinfo {author} {\bibfnamefont {R.}~\bibnamefont {Alléaume}},\ }\href@noop
  {} {\bibfield  {journal} {\bibinfo  {journal} {New J. of Phys.}\ }\textbf
  {\bibinfo {volume} {17}},\ \bibinfo {pages} {043027} (\bibinfo {year}
  {2015})}\BibitemShut {NoStop}%
\bibitem [{\citenamefont {McMahon}\ \emph {et~al.}(2014)\citenamefont
  {McMahon}, \citenamefont {Lund},\ and\ \citenamefont
  {Ralph}}]{McMahon_Optimal_QS}%
  \BibitemOpen
  \bibfield  {author} {\bibinfo {author} {\bibfnamefont {N.~A.}\ \bibnamefont
  {McMahon}}, \bibinfo {author} {\bibfnamefont {A.~P.}\ \bibnamefont {Lund}}, \
  and\ \bibinfo {author} {\bibfnamefont {T.~C.}\ \bibnamefont {Ralph}},\
  }\href@noop {} {\bibfield  {journal} {\bibinfo  {journal} {Phys. Rev. A}\
  }\textbf {\bibinfo {volume} {89}},\ \bibinfo {pages} {023846} (\bibinfo
  {year} {2014})}\BibitemShut {NoStop}%
\bibitem [{\citenamefont {Jeffers}(2010)}]{Jeffers_Two-photon_QS}%
  \BibitemOpen
  \bibfield  {author} {\bibinfo {author} {\bibfnamefont {J.}~\bibnamefont
  {Jeffers}},\ }\href@noop {} {\bibfield  {journal} {\bibinfo  {journal} {Phys.
  Rev. A}\ }\textbf {\bibinfo {volume} {82}},\ \bibinfo {pages} {063828}
  (\bibinfo {year} {2010})}\BibitemShut {NoStop}%
\bibitem [{\citenamefont {Navascu\'es}\ and\ \citenamefont
  {Ac\'{\i}n}(2005)}]{Navascues_EntanglingCloner}%
  \BibitemOpen
  \bibfield  {author} {\bibinfo {author} {\bibfnamefont {M.}~\bibnamefont
  {Navascu\'es}}\ and\ \bibinfo {author} {\bibfnamefont {A.}~\bibnamefont
  {Ac\'{\i}n}},\ }\href@noop {} {\bibfield  {journal} {\bibinfo  {journal}
  {Phys. Rev. Lett.}\ }\textbf {\bibinfo {volume} {94}},\ \bibinfo {pages}
  {020505} (\bibinfo {year} {2005})}\BibitemShut {NoStop}%
\bibitem [{\citenamefont {Nielsen}\ and\ \citenamefont
  {Chuang}(2000)}]{Nielsen_Chuang}%
  \BibitemOpen
  \bibfield  {author} {\bibinfo {author} {\bibfnamefont {M.~A.}\ \bibnamefont
  {Nielsen}}\ and\ \bibinfo {author} {\bibfnamefont {I.~L.}\ \bibnamefont
  {Chuang}},\ }\href@noop {} {\emph {\bibinfo {title} {Quantum Computation and
  Quantum Information}}}\ (\bibinfo  {publisher} {Cambridge University Press,
  Cambridge},\ \bibinfo {year} {2000})\BibitemShut {NoStop}%
\bibitem [{\citenamefont {Garc\'{\i}a-Patr\'on}\ and\ \citenamefont
  {Cerf}(2006)}]{Garcia-Patron_OptGaussAttacks}%
  \BibitemOpen
  \bibfield  {author} {\bibinfo {author} {\bibfnamefont {R.}~\bibnamefont
  {Garc\'{\i}a-Patr\'on}}\ and\ \bibinfo {author} {\bibfnamefont {N.~J.}\
  \bibnamefont {Cerf}},\ }\href@noop {} {\bibfield  {journal} {\bibinfo
  {journal} {Phys. Rev. Lett.}\ }\textbf {\bibinfo {volume} {97}},\ \bibinfo
  {pages} {190503} (\bibinfo {year} {2006})}\BibitemShut {NoStop}%
\bibitem [{\citenamefont {Pirandola}\ \emph {et~al.}(2008)\citenamefont
  {Pirandola}, \citenamefont {Braunstein},\ and\ \citenamefont
  {Lloyd}}]{Pirandola:PRL_GaussAtt2008}%
  \BibitemOpen
  \bibfield  {author} {\bibinfo {author} {\bibfnamefont {S.}~\bibnamefont
  {Pirandola}}, \bibinfo {author} {\bibfnamefont {S.~L.}\ \bibnamefont
  {Braunstein}}, \ and\ \bibinfo {author} {\bibfnamefont {S.}~\bibnamefont
  {Lloyd}},\ }\href@noop {} {\bibfield  {journal} {\bibinfo  {journal} {Phys.
  Rev. Lett.}\ }\textbf {\bibinfo {volume} {101}},\ \bibinfo {pages} {200504}
  (\bibinfo {year} {2008})}\BibitemShut {NoStop}%
\bibitem [{\citenamefont {{He}}\ \emph {et~al.}(2018)\citenamefont {{He}},
  \citenamefont {{Malaney}},\ and\ \citenamefont {{Green}}}]{Malaney_Sat}%
  \BibitemOpen
  \bibfield  {author} {\bibinfo {author} {\bibfnamefont {M.}~\bibnamefont
  {{He}}}, \bibinfo {author} {\bibfnamefont {R.}~\bibnamefont {{Malaney}}}, \
  and\ \bibinfo {author} {\bibfnamefont {J.}~\bibnamefont {{Green}}},\ }in\
  \href {\doibase 10.1109/GLOCOMW.2018.8644254} {\emph {\bibinfo {booktitle}
  {2018 IEEE Globecom Workshops (GC Wkshps)}}}\ (\bibinfo {year} {2018})\ pp.\
  \bibinfo {pages} {1--6}\BibitemShut {NoStop}%
\bibitem [{\citenamefont {Cover}\ and\ \citenamefont {Thomas}(2006)}]{Cover}%
  \BibitemOpen
  \bibfield  {author} {\bibinfo {author} {\bibfnamefont {T.~M.}\ \bibnamefont
  {Cover}}\ and\ \bibinfo {author} {\bibfnamefont {J.~A.}\ \bibnamefont
  {Thomas}},\ }\href@noop {} {\emph {\bibinfo {title} {Elements of Information
  Theory-Second Edition}}}\ (\bibinfo  {publisher} {John Wiley and Sons, New
  Jersey},\ \bibinfo {year} {2006})\BibitemShut {NoStop}%
\bibitem [{\citenamefont {Pirandola}\ \emph {et~al.}(2017)\citenamefont
  {Pirandola}, \citenamefont {Laurenza}, \citenamefont {Ottaviani},\ and\
  \citenamefont {Banchi}}]{Pirandola_PLOB17}%
  \BibitemOpen
  \bibfield  {author} {\bibinfo {author} {\bibfnamefont {S.}~\bibnamefont
  {Pirandola}}, \bibinfo {author} {\bibfnamefont {R.}~\bibnamefont {Laurenza}},
  \bibinfo {author} {\bibfnamefont {C.}~\bibnamefont {Ottaviani}}, \ and\
  \bibinfo {author} {\bibfnamefont {L.}~\bibnamefont {Banchi}},\ }\href@noop {}
  {\bibfield  {journal} {\bibinfo  {journal} {Nat. Commun.}\ }\textbf {\bibinfo
  {volume} {8}},\ \bibinfo {pages} {15043} (\bibinfo {year}
  {2017})}\BibitemShut {NoStop}%
\bibitem [{\citenamefont {M\"uller}\ \emph {et~al.}(2014)\citenamefont
  {M\"uller}, \citenamefont {Bounouar}, \citenamefont {J\"ons}, \citenamefont
  {Gl\"assl},\ and\ \citenamefont {Michler}}]{QDot_entg_low_g2:NatPhot2014}%
  \BibitemOpen
  \bibfield  {author} {\bibinfo {author} {\bibfnamefont {M.}~\bibnamefont
  {M\"uller}}, \bibinfo {author} {\bibfnamefont {S.}~\bibnamefont {Bounouar}},
  \bibinfo {author} {\bibfnamefont {K.~D.}\ \bibnamefont {J\"ons}}, \bibinfo
  {author} {\bibfnamefont {M.}~\bibnamefont {Gl\"assl}}, \ and\ \bibinfo
  {author} {\bibfnamefont {P.}~\bibnamefont {Michler}},\ }\href@noop {}
  {\bibfield  {journal} {\bibinfo  {journal} {Nat. Photon.}\ }\textbf {\bibinfo
  {volume} {8}},\ \bibinfo {pages} {224} (\bibinfo {year} {2014})}\BibitemShut
  {NoStop}%
\bibitem [{\citenamefont {Senellart}\ \emph {et~al.}(2017)\citenamefont
  {Senellart}, \citenamefont {Solomon},\ and\ \citenamefont
  {White}}]{Senellart_qdots:NatPhot2017}%
  \BibitemOpen
  \bibfield  {author} {\bibinfo {author} {\bibfnamefont {P.}~\bibnamefont
  {Senellart}}, \bibinfo {author} {\bibfnamefont {G.}~\bibnamefont {Solomon}},
  \ and\ \bibinfo {author} {\bibfnamefont {A.}~\bibnamefont {White}},\ }\href
  {\doibase 10.1038/nnano.2017.218} {\bibfield  {journal} {\bibinfo  {journal}
  {Nat. Nanotech.}\ }\textbf {\bibinfo {volume} {12}},\ \bibinfo {pages} {1026}
  (\bibinfo {year} {2017})}\BibitemShut {NoStop}%
\bibitem [{\citenamefont {Pirandola}\ \emph {et~al.}(2009)\citenamefont
  {Pirandola}, \citenamefont {Garc\'{\i}a-Patr\'on}, \citenamefont
  {Braunstein},\ and\ \citenamefont {Lloyd}}]{Pirandola:RevKeyCap2009}%
  \BibitemOpen
  \bibfield  {author} {\bibinfo {author} {\bibfnamefont {S.}~\bibnamefont
  {Pirandola}}, \bibinfo {author} {\bibfnamefont {R.}~\bibnamefont
  {Garc\'{\i}a-Patr\'on}}, \bibinfo {author} {\bibfnamefont {S.~L.}\
  \bibnamefont {Braunstein}}, \ and\ \bibinfo {author} {\bibfnamefont
  {S.}~\bibnamefont {Lloyd}},\ }\href@noop {} {\bibfield  {journal} {\bibinfo
  {journal} {Phys. Rev. Lett.}\ }\textbf {\bibinfo {volume} {102}},\ \bibinfo
  {pages} {050503} (\bibinfo {year} {2009})}\BibitemShut {NoStop}%
\bibitem [{\citenamefont {Garc\'{\i}a-Patr\'on}\ \emph
  {et~al.}(2009)\citenamefont {Garc\'{\i}a-Patr\'on}, \citenamefont
  {Pirandola}, \citenamefont {Lloyd},\ and\ \citenamefont
  {Shapiro}}]{Garcia-Patron:RevCohInf2009}%
  \BibitemOpen
  \bibfield  {author} {\bibinfo {author} {\bibfnamefont {R.}~\bibnamefont
  {Garc\'{\i}a-Patr\'on}}, \bibinfo {author} {\bibfnamefont {S.}~\bibnamefont
  {Pirandola}}, \bibinfo {author} {\bibfnamefont {S.}~\bibnamefont {Lloyd}}, \
  and\ \bibinfo {author} {\bibfnamefont {J.~H.}\ \bibnamefont {Shapiro}},\
  }\href@noop {} {\bibfield  {journal} {\bibinfo  {journal} {Phys. Rev. Lett.}\
  }\textbf {\bibinfo {volume} {102}},\ \bibinfo {pages} {210501} (\bibinfo
  {year} {2009})}\BibitemShut {NoStop}%
\bibitem [{\citenamefont {Garc\'{\i}a-Patr\'on}\ and\ \citenamefont
  {Cerf}(2009)}]{PRL_trustednoise1}%
  \BibitemOpen
  \bibfield  {author} {\bibinfo {author} {\bibfnamefont {R.}~\bibnamefont
  {Garc\'{\i}a-Patr\'on}}\ and\ \bibinfo {author} {\bibfnamefont {N.~J.}\
  \bibnamefont {Cerf}},\ }\href@noop {} {\bibfield  {journal} {\bibinfo
  {journal} {Phys. Rev. Lett.}\ }\textbf {\bibinfo {volume} {102}},\ \bibinfo
  {pages} {130501} (\bibinfo {year} {2009})}\BibitemShut {NoStop}%
\bibitem [{\citenamefont {Madsen}\ \emph {et~al.}(2012)\citenamefont {Madsen},
  \citenamefont {Usenko}, \citenamefont {Lassen}, \citenamefont {Filip},\ and\
  \citenamefont {Andersen}}]{NCOM_trustednoise2}%
  \BibitemOpen
  \bibfield  {author} {\bibinfo {author} {\bibfnamefont {L.~S.}\ \bibnamefont
  {Madsen}}, \bibinfo {author} {\bibfnamefont {V.~C.}\ \bibnamefont {Usenko}},
  \bibinfo {author} {\bibfnamefont {M.}~\bibnamefont {Lassen}}, \bibinfo
  {author} {\bibfnamefont {R.}~\bibnamefont {Filip}}, \ and\ \bibinfo {author}
  {\bibfnamefont {U.~L.}\ \bibnamefont {Andersen}},\ }\href {\doibase
  10.1038/ncomms2097} {\bibfield  {journal} {\bibinfo  {journal} {Nature
  Communications}\ }\textbf {\bibinfo {volume} {3}},\ \bibinfo {pages} {1083}
  (\bibinfo {year} {2012})}\BibitemShut {NoStop}%
\end{thebibliography}%
\bibliographystyle{apsrev4-1}
%%%%%%%%%%%%%%%%%%%%%%%%%%%%%%%%%%%%%

\appendix 

 \section{Conditional output state $\hat{\omega}_{\mathrm{out}}^{\mathrm{PS}}(x_A) $}
 \label{app:output_dist_cond}
In order to find the conditional output state when Alice has used an $X$ quadrature 
value of $x_A$, we start with the input state in \eqref{input_st_QS}, and take an average over $P_A$ with the input Gaussian distribution of $f_{P_A}(p_A) = e^{-\frac{p_A^2}{V_A/2}}/\sqrt{\pi V_A/2}$. As a result, the output characteristic function in \eqref{A-func-out}  will also be averaged out and result in the following output state:
 \begin{align}
 	\hat{\omega}_{\mathrm{out}}^{\mathrm{PS}} (x_A)= & \omega_{00}(x_A) |0\rangle_{\hat{b}_3}\langle 0| +  \omega_{01}(x_A) |0\rangle_{\hat{b}_3}\langle 1| \nonumber \\
 	& + \omega_{10}(x_A) |1\rangle_{\hat{b}_3}\langle 0| + \omega_{11}(x_A)  |1\rangle_{\hat{b}_3}\langle 1|  ,
 \end{align}
 where
 \begin{align}
 	\begin{cases}
 		\omega_{00}(x_A)= \frac{\widetilde{\omega}_{00}(x_A)}{P^{\rm PS}(x_A)}  \\
 		\omega_{01}(x_A) = \omega_{10}^\ast(x_A) =  \frac{\widetilde{\omega}_{01}(x_A)}{P^{\rm PS}(x_A)}   \\
 		\omega_{11}(x_A)= \frac{\widetilde{\omega}_{11}(x_A)}{P^{\rm PS}(x_A)}  ,            
 	\end{cases}
 \end{align}
 with 
 \begin{align}
 	\begin{cases}
 		\widetilde{\omega}_{00}(x_A)= \frac{ 8F_1(2F_1+1)^2 + TV_A(8F_1^2+6F_1+1) + 2T(TV_A+4F_1+2)x_A^2 }{ (g^2+1) (2F_1+1)^{5/2} (TV_A+4F_1+2)^{3/2} } \\
 		~~~~~~~~~~~~~ \times \sqrt{2} e^{-\frac{Tx_A^2}{2F_1+1} } \\
 		\widetilde{\omega}_{01}(x_A)= -\frac{2g\sqrt{2T}x_A}{(g^2+1) (2F_1+1)^{3/2} \sqrt{TV_A+4F_1+2} }  e^{-\frac{Tx_A^2}{2F_1+1} } \\
 		\widetilde{\omega}_{11}(x_A)=  \frac{g^2}{g^2+1} \Big(  \frac{2\sqrt{2}e^{-\frac{Tx_A^2}{2F_1+1}}  }{ \sqrt{(2F_1+1) (TV_A+4F_1+2)} } -  \frac{e^{-\frac{Tx_A^2}{2F_1}} }{\sqrt{F_1(TV_A+4F_1)} }   \Big)  \\
 		P^{\rm PS}(x_A) =\widetilde{\omega}_{00}(x_A)+\widetilde{\omega}_{11}(x_A).
 	\end{cases} \nonumber 
 \end{align}
 
 \section{Covariance matrix elements}
 \label{app:CM_elements}
 Having obtained the output antinormally-ordered characteristic function of \eqref{EPRout-anti-nor}, we use \eqref{rec-state} to find the corresponding output state:
 \begin{align}
 	\hat{\rho}_{0123\mathrm{N}}^{\mathrm{out}} & = \int \frac{d^2\xi_0}{\pi}\frac{d^2\xi_1}{\pi}\frac{d^2\xi_2}{\pi}\frac{d^2\xi_3}{\pi}\frac{d^2\xi_\mathrm{N}}{\pi}  \chi_{\mathrm{A}}^{\mathrm{out}}(\xi_0, \xi_1,\xi_2, \xi_3 ,\xi_\mathrm{N}) \nonumber \\
 	& \hat{D}_{\mathrm{N}}(\hat{a}_0,\xi_0) \hat{D}_{\mathrm{N}}(\hat{b}_1,\xi_1) \hat{D}_{\mathrm{N}}(\hat{b}_2,\xi_2) \hat{D}_{\mathrm{N}}(\hat{b}_3,\xi_3) \hat{D}_{\mathrm{N}}(\hat{b}_\mathrm{N},\xi_\mathrm{N}). \nonumber 
 \end{align}
 In the following, we show how the shared state between Alice and Bob is found step-by-step. 
 We first trace out mode $\hat{b}_\mathrm{N}$, see Fig.~\ref{fig:EPRQS}, to obtain
 \begin{align}
 	\hat{\rho}_{0123}^{\mathrm{out}}= & \int \frac{d^2\xi_0}{\pi}\frac{d^2\xi_1}{\pi}\frac{d^2\xi_2}{\pi}\frac{d^2\xi_3}{\pi}  \chi_{\mathrm{A}}^{\mathrm{out}}(\xi_0, \xi_1,\xi_2, \xi_3 ,0)  \nonumber \\
 	& \hat{D}_{\mathrm{N}}(\hat{a}_0,\xi_0) \hat{D}_{\mathrm{N}}(\hat{b}_1,\xi_1) \hat{D}_{\mathrm{N}}(\hat{b}_2,\xi_2) \hat{D}_{\mathrm{N}}(\hat{b}_3,\xi_3),
 \end{align}
 where we used  $\mathrm{tr}[\hat{D}_{\mathrm{N}}(a,\xi)]=\pi \delta^2(\xi)$. Next, by defining the measurement operator $ \hat{M} = (\mathbbm{1} -|0\rangle_{b_1}\langle 0|)\otimes |0\rangle_{b_2}\langle 0|$, modes $\hat{b}_1$ and $\hat{b}_2$ are measured. The post-selected state is
 \begin{align}
 	\hat{\rho}_{03}^{\mathrm{PS}}= \frac{ \mathrm{tr}_{12}( \hat{\rho}_{0123}^{\mathrm{out}} \hat{M} ) }{ \mathrm{tr} ( \hat{\rho}_{0123}^{\mathrm{out}} \hat{M} ) } =: \frac{\hat{\sigma}_{03}^{\mathrm{PS}}}{P_{\mathrm{EB}}^{\mathrm{PS}} },
 \end{align}
 where 
 \begin{align}
 	\hat{\sigma}_{03}^{\mathrm{PS}}=  \int \frac{d^2\xi_0}{\pi}\frac{d^2\xi_3}{\pi}  \widetilde{\chi}_{\mathrm{A}}(\xi_0,\xi_3)  \hat{D}_{\mathrm{N}}(\hat{a}_0,\xi_0)  \hat{D}_{\mathrm{N}}(\hat{b}_3,\xi_3)  
 \end{align}
 with 
 \begin{align}
 	\label{chibarfunc}
 	\widetilde{\chi}_{\mathrm{A}}(\xi_0,\xi_3) =   \int   
 	\frac{d^2\xi_1}{\pi}\frac{d^2\xi_2}{\pi} \chi_{\mathrm{A}}^{\mathrm{out}}(\xi_0, \xi_1,\xi_2, \xi_3 ,0) \big(\pi \delta^2(\xi_1)-1   \big) , 
 \end{align}
 and $P_{\rm EB}^{\mathrm{PS}} = \overline P_{\rm succ}/2$ is the corresponding success probability to measurement $\hat M$:  
 \begin{align}
 	P_{\mathrm{EB}}^{\mathrm{PS}} = & \int \frac{d^2\xi_1}{\pi}\frac{d^2\xi_2}{\pi} \chi_{\mathrm{A}}^{\mathrm{out}}(0, \xi_1, \xi_2, 0 ,0) \big(\pi \delta^2(\xi_1)-1   \big)  \nonumber \\
 	= & \widetilde{\chi}_{\mathrm{A}}(0,0).
 \end{align}
 
Now, we find the CM for $\hat{\rho}_{03}^{\mathrm{PS}}$. In doing so, we need to work out the triplet $(a,b,c)$ of the corresponding CM as follows. By definition, assuming that $\hat{x}_{0}$ is the $X$ quadrature of mode $\hat{a}_0$, we have
 \begin{align}
 	a= \langle \hat{x}_{0}^2 \rangle_{\hat{\rho}_{03}}  = & \frac{\langle \hat{x}_{0}^2 \rangle_{\hat{\sigma}_{03}} }{P_{\mathrm{EB}}^{\mathrm{PS}}} = \frac{\mathrm{tr}(\hat{\sigma}_{03} \hat{x}_{0}^2 )}{ P_{\mathrm{EB}}^{\mathrm{PS}} },
 \end{align}
 where
 \begin{align}
 	\mathrm{tr}(\hat{\sigma}_{03} \hat{x}_{0}^2) = & \int \frac{d^2\xi_0}{\pi}\frac{d^2\xi_3}{\pi}  \widetilde{\chi}_{\mathrm{A}}(\xi_0,\xi_3) \nonumber \\
 	& \times \mathrm{tr}[\hat{x}_{0}^2 \hat{D}_{\mathrm{N}}(\hat{a}_0,\xi_0) ] \times  \mathrm{tr}[ \hat{D}_{\mathrm{N}}(\hat{b}_3,\xi_3) ]  \nonumber \\
 	= &  \int \frac{d^2\xi_0}{\pi}   \widetilde{\chi}_{\mathrm{A}}(\xi_0,0)  \times  \mathrm{tr}( \hat{D}_{\mathrm{N}}(\hat{a}_0,\xi_0) \hat{x}_{0}^2) .
 \end{align}
 Assuming that $\xi_0=x+iy$, one can show that $\mathrm{tr}( \hat{D}_{\mathrm{N}}(\hat{a}_0,\xi_0) \hat{x}_{0}^2 )=\pi \delta^2(\xi_0)+ 2\pi y \delta(x) \frac{d}{dy}\delta(y) -\pi \delta(x)\frac{d^2}{dy^2}\delta(y)$; thus, 
 \begin{align}
 	\mathrm{tr}(\hat{\sigma}_{03} \hat{x}_{0}^2 ) = & -\widetilde{\chi}_{\mathrm{A}}(0,0)  - \frac{d^2}{dy^2}  \widetilde{\chi}_{\mathrm{A}}(0,y,\xi_3=0) \Big|_{y=0},
 \end{align}
 where we use the identity $\int dz f(z) \frac{d}{dz}\delta(z)= - \int dz \frac{d}{dz} f(z) \delta(z) $. Therefore, 
 \begin{align}
 	a= -1-\frac{\frac{d^2}{dy^2}  \widetilde{\chi}_{\mathrm{A}}(0,y,\xi_3=0) \Big|_{y=0}}{\widetilde{\chi}_{\mathrm{A}}(0,0)}.
 \end{align}
 In a similar way, assuming $\xi_0=x+iy$ and $\xi_3=u+iv$, we show that
 \begin{align}
 	b=\frac{\mathrm{tr}( \hat{\sigma}_{03} \hat{x}_{3}^2 )}{\widetilde{\chi}_{\mathrm{A}}(0,0) }   = -1-\frac{\frac{d^2}{dv^2}  \widetilde{\chi}_{\mathrm{A}}(\xi_0=0,0,v) \Big|_{v=0}}{\widetilde{\chi}_{\mathrm{A}}(0,0)} 
 \end{align}
 and
 \begin{align}
 	c= \frac{\mathrm{tr}( \hat{\sigma}_{03}\hat{x}_{0} \hat{x}_{3} ) }{\widetilde{\chi}_{\mathrm{A}}(0,0) } =\frac{ \frac{d}{dv} \Big[\frac{d}{dy}  \widetilde{\chi}_{\mathrm{A}}(0,y,0,v) \Big|_{y=0}\Big]\Big|_{v=0}  }{ \widetilde{\chi}_{\mathrm{A}}(0,0)}.
 \end{align}
 Having the integrals in \eqref{chibarfunc} taken, we are able to calculate the triplet $(a,b,c)$, thus the CM. 
 Using \textsc{maple}, we obtain the closed form expressions as summarized in \eqref{abc}.  
 
 Having the triplet $(a,b,c)$, $\chi_{\rm BE}^\star$ is upper bounded by:
 \begin{align}
 	\label{app:Eq-Holevo}
 	\chi_{\rm BE}^{\rm G} = g(\Lambda_1) + g(\Lambda_2)  -  g(\Lambda_3), 
 \end{align}
 where 
 $$g(x)= (\frac{x+1}{2}) \log_2             (\frac{x+1}{2})  - (\frac{x-1}{2})  \log_2 (\frac{x-1}{2})  $$
 and $ \Lambda_{1/2} = \sqrt{( A \pm \sqrt{A^2 - 4 B^2})/2 }   =(\sqrt{(a+b)^2-4 c^2}\pm (b-a))/2 , ~~~\Lambda_{3} =\sqrt{a B/b}=\sqrt{a( ab- c^2)/b}  $, with $A= a^2 + b^2 - 2 c^2$ and $B= ab- c^2$. Note that \eqref{app:Eq-Holevo} is valid when we neglect the electronic noise at the receiver as we have assumed in our numerical results.     
 Also, mutual information can be calculated form the covariance matrix, if we wish to use the Gaussian approximation, by
 \begin{align}
 	\label{Eq:IABG}
 	I_{\rm AB}^{\rm G}= \frac{1}{2} \log_2 \frac{ab}{ab-c^2}.
 \end{align}

\end{document}